\documentclass{aa}

\usepackage{graphicx}
\usepackage{txfonts}
\usepackage{xcolor}
\usepackage{ulem}
\usepackage{multicol}
\usepackage[toc]{appendix}
\newcommand{\exafe}{{{ex$\alpha$fe~}}}
\newcommand{\exafey}{{{y-ex$\alpha$fe~}}}

\newcommand{\afe}{[$\alpha$/Fe]~}
\newcommand{\feh}{[Fe/H]~}
\newcommand{\ms}{M$_{\odot}$}
\newcommand{\anormal}{[$\alpha$/Fe]-normal}

\begin{document} 

   \title{Lithium, masses, and kinematics of young Galactic dwarf and giant stars with extreme \afe ratios}

   \author{S.~Borisov
          \inst{1}
          \and
          N.~Prantzos\inst{2}
          \and
          C.~Charbonnel\inst{1,3}
          }

   \institute{Department of Astronomy, University of Geneva, Chemin Pegasi 51, 1290 Versoix, Switzerland\\
              \email{sviatoslav.borisov@unige.ch}
         \and
            Institut d’Astrophysique de Paris, UMR7095 CNRS, Sorbonne Université, 98bis Bd. Arago, 75104 Paris, France
         \and
            IRAP, CNRS UMR 5277 \& Université de Toulouse, 14 avenue Edouard Belin, 31400 Toulouse, France
             }

  \abstract
   {Recent spectroscopic explorations of large Galactic stellar samples stars have revealed the existence of red giants with \afe ratios that are anomalously high, given their relatively young ages.}
   {We revisit the GALAH~DR3 survey to look for both dwarf and giant stars with 
   extreme \afe ratios, that is, the upper 1\% in the [$\alpha$/Fe]-[Fe/H] plane  over the range in [Fe/H] between -1.1 and +0.4~dex. We refer to these outliers as ``ex$\alpha$fe'' stars.} 
   {We used the GALAH~DR3 data along with their value-added catalog to trace the properties (chemical abundances, masses, ages, and kinematics) of the {ex$\alpha$fe stars}. We applied strict criteria to the quality of the determination of the stellar parameters, abundances, and age determinations to select our sample of single stars. We investigated the effects of secular stellar evolution and the magnitude limitations of the GALAH survey to understand the mass and metallicity distributions of the sample stars. Here, we also discuss the corresponding biases in previous studies of stars with high -- albeit not extreme -- \afe in other spectroscopic surveys.}
   {We find both dwarf and giant \exafe stars younger than 3 Gyr, which we refer to as ``y-ex$\alpha$fe'' stars. Dwarf \exafey stars exhibit lithium abundances similar to those of young \anormal~dwarfs at the same age and [Fe/H]. In particular, the youngest and most massive stars of both populations exhibit the highest Li abundances, A(Li)$\sim$3.5~dex (i.e., a factor of 2 above the protosolar value), while cooler and older stars exhibit the same Li depletion patterns increasing with both decreasing mass and increasing age. In addition, the [Fe/H] and mass distributions of both the dwarf and giant \exafey stars do not differ from those of their \anormal~counterparts found in the thin disk and they share the same kinematic properties, with lower eccentricities and velocities with respect to the local standard of rest than old stars of the thick disk.} 
  {We conclude that \exafey dwarf and giant stars are indeed young, their mass distribution shows no peculiarity, and they differ from young \anormal~stars by their extreme \afe content only.  However, their origins still remain unclear.}

   \maketitle
%
\section{Introduction}
\label{section:intro}

One of the key questions in modern astrophysics concerns the formation and evolution of the Milky Way and its different substructures. Our Galaxy is a complex system consisting of different components 
\citep[e.g.,][and references therein]{HelmiARAA2020,2022arXiv220605534G}, with the most prominent ones in the solar neighborhood being the thin and thick disks and the halo. The various aspects of the formation and subsequent dynamical and chemical evolution of each of those components, as well as the connections between them, are still poorly understood  \citep[e.g.,][and references therein]{BlandHawthorn2016ARA&A}. 

Since the discovery of the thick disk \citep{Gilmore1983}, it has been established that this component differs from the thin disk not only in terms of its spatial properties (i.e.,  it is more extended vertically), but also the kinematic properties \citep[i.e., lower rotational velocity and higher velocity dispersion, e.g.,][]{Chiba2000}, as well as in the corresponding stellar ages, which are higher for the thick disk \citep[e.g.][]{Reddy2006}. Regarding the chemical properties, the thick disk has lower metallicity (on average) and higher \afe ratios than the thin disk for the same metallicity \citep[e.g.,][]{Prochaska2000,Mishenina2004}. This latter property is understood as a natural consequence of the higher age of the thick disk which is supposed to evolve in a short timescale (a few Gy); this leaves less time to SN~Ia to enrich the interstellar medium with their Fe-rich ejecta, in contrast to the thin disk which evolves on longer timescales of many Gyr. However, the observed co-existence of two clearly distinguished quasi-parallel \afe sequences in the same region of space -- the solar cylinder -- does not yet have a commonly accepted interpretation: very broadly, this could be due either to some specific dynamical event, such as an early merger or ``quenching'' of star formation followed by secondary infall \citep{Larson1980} or by secular disk evolution \citep[e.g.,][]{Schoenrich2009,Loebman2011}.

\citet{adibekyan2011} found that high \afe dwarf stars appear separated into two families with a gap in the distributions of both \afe and metallicity at [$\alpha$/Fe]$\sim$0.17 and [Fe/H]$\sim-0.2)$. They found that both the metal-poor high-\afe stars (thick disk) and the metal-rich high-\afe stars are, on average, older than the chemically defined thin disk stars (low-\afe stars). They adopted the term ``h$\alpha$mr'' to characterize high-\afe and metal-rich stars, which were found to have kinematics and orbits similar to the thin disk stars, but are older by a few Gyr, that is, they have ages intermediate between those of the thick and thin disks.

The use of large spectroscopic surveys, such as RAVE \citep{Steinmetz2006,Steinmetz2020}, APOGEE \citep{Majewski2017}, LAMOST \citep{Zhao2012,Luo2015}, GALAH \citep{Buder2021} or Gaia-ESO survey \citep{Gilmore2012},  have provided  a wealth of opportunities  in the fields of stellar physics and Galactic archaeology over the past few years.
A number of recent studies have reported  young giant stars with \afe values higher than predicted by standard chemical evolution models of the Milky Way \citep{Chiappini2015,Martig2015,2018MNRAS.475.5487S,2018MNRAS.475.3633W,Miglio2020,Zhang2021}. Using CoRot \citep{Baglin2006} and APOGEE data, \citet{Chiappini2015} found that these stars have a lower iron-peak element content than the rest of the sample and are more abundant towards the inner Galactic disk regions. Their tentative interpretation of these observations is that these stars were formed close to the end of the Galactic bar, that is, near corotation. This is a region where gas can be kept inert for longer times than in other regions that are more frequently shocked by the passage of spiral arms and where the mass return from older inner-disk stellar generations is expected to be highest (according to an inside-out disk-formation scenario), which additionally dilutes the in-situ gas. On the other hand, using LAMOST and Gaia~DR2, \citet{Zhang2021} found similar red giant stars with high masses and sharing the same kinematics as the high-\afe old stellar population in the Galactic thick disk. According to \citet{Zhang2021}, these stars mimic ``young'' single stars, but they actually belong to an intrinsic old stellar population, as the thick disk. Similarly, \citet{Miglio2020}, studying red-giant stars with exquisite asteroseismic (Kepler), spectroscopic (APOGEE), and astrometric (Gaia) data find that massive (M$\geq$1.1~M$_{\odot}$) [$\alpha$/Fe]-rich stars are a fraction of $\sim$5\% on the RGB, and significantly higher in the red clump, apparently supporting the scenario according to which most of these stars had undergone an interaction with a companion. 

In this paper, we focus on the uppermost envelope (the upper 1\%) of the \afe values of dwarf stars of GALAH, independently of their metallicity. We show that some of them have unexpectedly young ages. We propose to use the term \exafe for those stars with extremely high \afe values, and \exafey for the youngest of them (i.e., with ages below 3~Gyr, see below). We begin in \S~\ref{high_alpha_young_dwarfs} with an overview of our dwarf sample and the relevant indicators of age for \exafey stars. In \S~\ref{high_alpha_giants}, we describe the young [$\alpha$/Fe]-rich stars problem for giants and compare the results obtained for giants with the use of the LAMOST survey and other selection criteria. Finally, we summarize our results and present our conclusions in \S~\ref{sec:summary}.


\section{Extreme [$\alpha$/Fe]-rich ({{ex$\alpha$fe}}) dwarf stars}
\label{high_alpha_young_dwarfs}

\begin{table*}
\centering
\begin{tabular}{|l|ccc|ccc|} 
\hline
&Main & \exafe & \exafey &Additional & add-\exafe & add-\exafey \\
Criterion          &   &   &   & &   &  \\
\hline
\hline
log~$g$; $T_\mathrm{eff}$&$\geq$3.5; $\geq$ 5500 & $\geq$3.5;  $>$ 5500 & $\geq$3.5; $\geq$ 5500 & $\geq$3.5; $\geq$ 5500 & $\geq$3.5; $\geq$ 5500 & $\geq$3.5; $\geq$ 5500\\\
      & $\geq$3.8; $<$ 5500          & $\geq$3.8; $<$ 5500         & $\geq$3.8; $<$ 5500         & $\geq$3.8; $<$ 5500 & $\geq$3.8; $<$ 5500 & $\geq$3.8; $<$ 5500 \\\
[Fe/H]          &   -1.1$\div$0.4~dex   & -1.1$\div$0.4~dex & -1.1$\div$0.4~dex & -1.1$\div$0.4~dex & -1.1$\div$0.4~dex & -1.1$\div$0.4~dex\\\
[$\alpha$/Fe]   &   -0.15$\div$0.7~dex  & top 1\%  of Main         &   top 1\% of Main        & -0.15$\div$0.7~dex &  top 1\% of Add  &  top 1\% of Add \\
$\sigma_{[\alpha/\mathrm{Fe}]}$   & $\leq0.05$~dex  & $\leq0.05$~dex & $\leq0.05$~dex & $\leq0.05$~dex & $\leq0.05$~dex & $\leq0.05$~dex\\
Age &   0.5$\div$13~Gyr &   0.5$\div$13~Gyr & 0.5$\div$3~Gyr  & 0.5$\div$13~Gyr & 0.5$\div$13~Gyr & 0.5$\div$3~Gyr\\
$\sigma_\mathrm{age}/\mathrm{age}$ & $\leq$30\%    & $\leq$30\% & $\leq$30\% & $>$30\% & $>$30\% & $>$30\% \\
S/N & $\geq$30   & $\geq$30 & $\geq$30    & $\geq$30 & $\geq$30 & $\geq$30 \\
{\small{\texttt{flag\_sp}}} & +    & + & + & + & + & + \\
{\small{\texttt{flag\_fe\_h}}} & +    & + & +  & + & + & +\\
{\small{\texttt{flag\_alpha\_fe}}} & +    & + & +  & + & + & +\\
{\small{\texttt{flag\_li\_fe}}}   & -- (+)             & -- (+)       & -- (+)    & -- (+) & -- (+) & -- (+) \\
A(Li)               & --        & --  & -- & --  & --  & -- \\\
    & ($\geq$0)      & ($\geq$0)  &  ($\geq$0)   & ($\geq$0) & ($\geq$0)  &  ($\geq$0) \\
\hline
\hline
Total number & 105~297     &  1~122  & 280 & 33~947  & 632 & 240  \\
 & (55~508) &  (222) &  (55)  & (11~326)  &  (74)  &  (38)  \\
\hline
\end{tabular}
\caption{Selection criteria applied to build the different sub-samples of dwarfs we consider in this work. Numbers in parenthesis indicate the number of objects when both Li criteria (\texttt{flag\_li\_fe}=0 and A(Li)$\geq$0) are applied. }
\label{tab_sample}
\end{table*}

\subsection{Selection of the dwarf sample(s)}
\label{sample_selection}

\begin{figure}
    \center
     \includegraphics[width=1\linewidth]{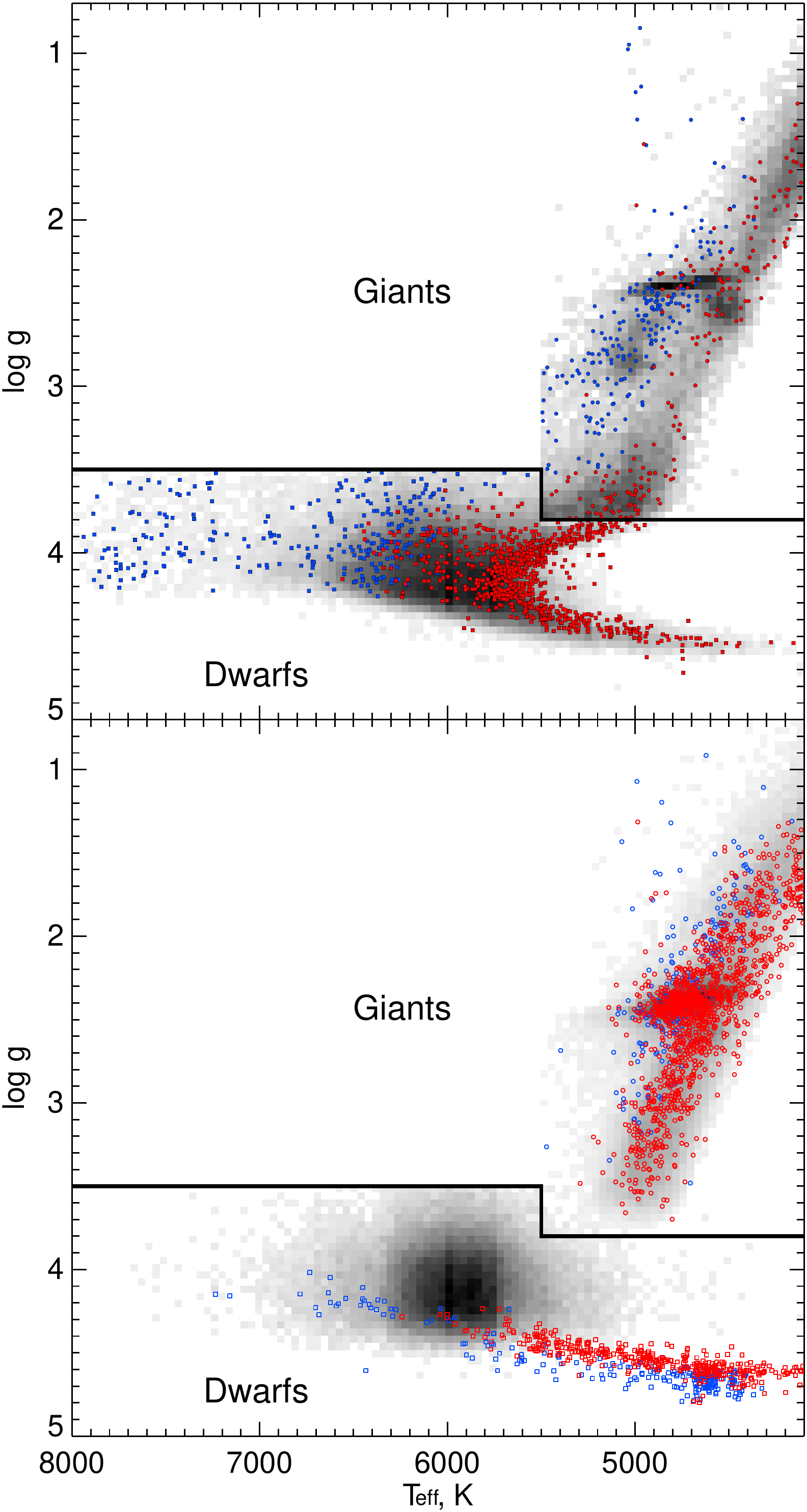}
     \caption{Kiel diagram for the dwarf and giant stars selected from GALAH~DR3. {\it Upper panel}: Number density plot showing the positions of $\alpha$-normal stars of the main samples of dwarfs and giants. Squares show the positions of extreme [$\alpha$/Fe]-rich ({{ex$\alpha$fe}}, see Sec.~\ref{definitionalpharich}) dwarf stars: \exafey (age$\leq$3~Gyr) and older \exafe are shown by blue and red squares, respectively. We keep the same notations, but in circles, for giants. {\it Bottom panel}: Same details, but for stars in the additional sample (open symbols).}
     \label{kiel_diagram}
 \end{figure}

The selection criteria for the dwarf samples are described below and summarized in Table~\ref{tab_sample}. The locations of the selected stars in the Kiel diagram are shown in Fig.~\ref{kiel_diagram}.

\subsubsection{Stellar parameters and abundances}
\label{stellar_parameters_abundances}

We selected dwarf stars with -1.1$\leq$[Fe/H]$\leq$0.4~dex and in a wide [$\alpha$/Fe] range (-0.15$\leq$[$\alpha$/Fe]$\leq$0.7) with ages and masses from the third data release of the GALAH survey, namely, the value-added catalog \citep[][VAC]{Buder2021}. 
We eliminated the binaries and pre-main sequence stars (\texttt{flag\_sp}=0). We selected stars with log~$g\geq$3.5 when $T_\mathrm{eff}$~$\geq$~5500K and log~$g\geq$3.8 when $T_\mathrm{eff}$~<~5500K, as shown in Fig.~\ref{kiel_diagram}; thus focusing our attention on dwarfs and possible subgiants to avoid evolved stars with Li abundances modified by the first dredge-up (e.g., \citealt{2020A&A...633A..34C,Martell2021}).
We applied strict quality criteria to assure a reliable determination of the stellar parameters ($T_\mathrm{eff}$, log~$g$) and of the values of [Fe/H] and [$\alpha$/Fe] (i.e., we use the following flags: \texttt{flag\_sp}=0, \texttt{flag\_fe\_h}=0, \texttt{flag\_alpha\_fe}=0, and S/N per pixel $\geq$ 30).

The [$\alpha$/Fe] values reported in the GALAH~DR3 catalog are calculated as an error-weighted combination of selected individual lines of Mg, Si, Ca, and Ti (nine lines in total for all the four elements). In this combination, the abundances of three elements are non-LTE: Mg \citep{Osorio2015}, Si \citep{Amarsi2017}, and Ca \citep{Osorio2019}. If one or more of the selected lines ``fail'' the quality test, the estimation is a combination of the rest, but no information is given on the actual number of lines and the actual combination of Mg, Si, Ca, and Ti that were used to compute the [$\alpha$/Fe] values for individual stars. Since we place special emphasis on the content of $\alpha$-elements in young stars, we selected those objects with $\sigma_{[\alpha/\mathrm{Fe}]}\leq0.05$~dex. 

Finally, to study the Li behavior of the dwarf stars, we selected among them those with reliable [Li/Fe] (\texttt{flag\_li\_fe}=0 in VAC). We computed A(Li)=[Li/Fe]+[Fe/H]+A(Li)$_{\odot}$ with A(Li)$_{\odot}$=0.96~dex \citep{Wang2021} and considered only those objects with A(Li)$\geq$0~dex.

\subsubsection{Ages, masses, and corresponding uncertainties}
\label{ages_sample}

In this work, we consider the ages and masses provided in GALAH~VAC that were obtained with the Bayesian stellar parameter estimation code BSTEP \citep{Sharma2018} using \texttt{PARSEC+COLIBRI} isochrones from \citet{Marigo2017}. However, to better handle the uncertainties on these two quantities, we independently computed the ages and masses of the dwarf sample with the SPInS tool \citep{lebreton2020} which relies on pre-computed stellar models (we used the solar-scaled BaSTI evolutionary tracks from \citealt{Pietrinferni2004,Pietrinferni2006})
and MCMC approach. 
The input observables used for the stars are $T_\mathrm{eff}$, luminosity $L$, log~$g$, and [M/H] (we computed [M/H] from [Fe/H] and [$\alpha$/Fe] using the formula from \citealt{Salaris2005}). We computed the luminosity using  Gaia~EDR3 $G$-band photometry based on the formula $L/L_{\odot}=10^{0.4\cdot(M_{\odot}^{bol}-(G+5-5\log(d)+BC_G-A_G))}$, where $M_{\odot}^{bol}=4.74^m$ \citep{prsa2016}, $BC_G$ is the bolometric correction computed according to \citet{andrae2018}, $d$ is the distance \citep{bailer-jones2020}, and $A_G$ is interstellar extinction. The mean relative difference between ``GALAH ages'' and ``SPInS ages'' is 32\%, with the mean relative uncertainties $\sigma$ on age of $\sim$30\% and $\sim$37\% from GALAH and SPInS, respectively. 
While the age uncertainties are quite large, which still remains one of the problems in stellar astrophysics, this comparison confirms that the vast majority of the stars that are young according to GALAH are indeed young in terms of single-stellar-evolution scenario -- independently of the methods based on isochrones or evolutionary tracks. 

In addition to the criteria described in \S~\ref{stellar_parameters_abundances}, we  only kept stars with relative uncertainties of age lower than its mean uncertainty ($\sigma_\mathrm{age}/\mathrm{age}\leq$30\%) from VAC for what we consider as our main sample. This is the sample with the strictest criteria for all the stellar properties that we discuss in this paper. On the other hand, isochrone-based age determination is (per se) less accurate in some areas of the Hertzsprung-Russel diagram (or Kiel diagram), in particular near the zero age main sequence (ZAMS) for the dwarfs with the lowest masses that exhibit both the longest MS lifetimes and the shortest MS ``paths'' in the HRD. Thus, we also included a separate sample with stars with higher age uncertainties ($\sigma_\mathrm{age}/\mathrm{age}>$30\%). This ``additional''\ sample contains mostly low-mass stars that lie close to the ZAMS and which are nearly absent in the sample (Fig.~\ref{kiel_diagram}). In both the main and additional samples, we eliminate stars with ages $<$0.5~Gyr and $>$13~Gyr because the age distribution shows sharp peaks in these ranges, which is probably caused by the convergence of the method to the lower and upper limits of the stellar evolution models grid. The resulting main and additional samples contain 105~297 and 33~947 dwarf stars, respectively. The main and additional ``Li-sub-samples'' (stars with measured Li abundances) contain 55~508 and 11~326 dwarf stars, respectively (Table~\ref{tab_sample}). 

Regarding the stellar masses, we put no constraint on the mass uncertainty because the mean mass uncertainties for our dwarf sample is low ($\sim$4\%).

\begin{figure*}
    \center
    \begin{multicols}{2}
    \includegraphics[width=1\linewidth]{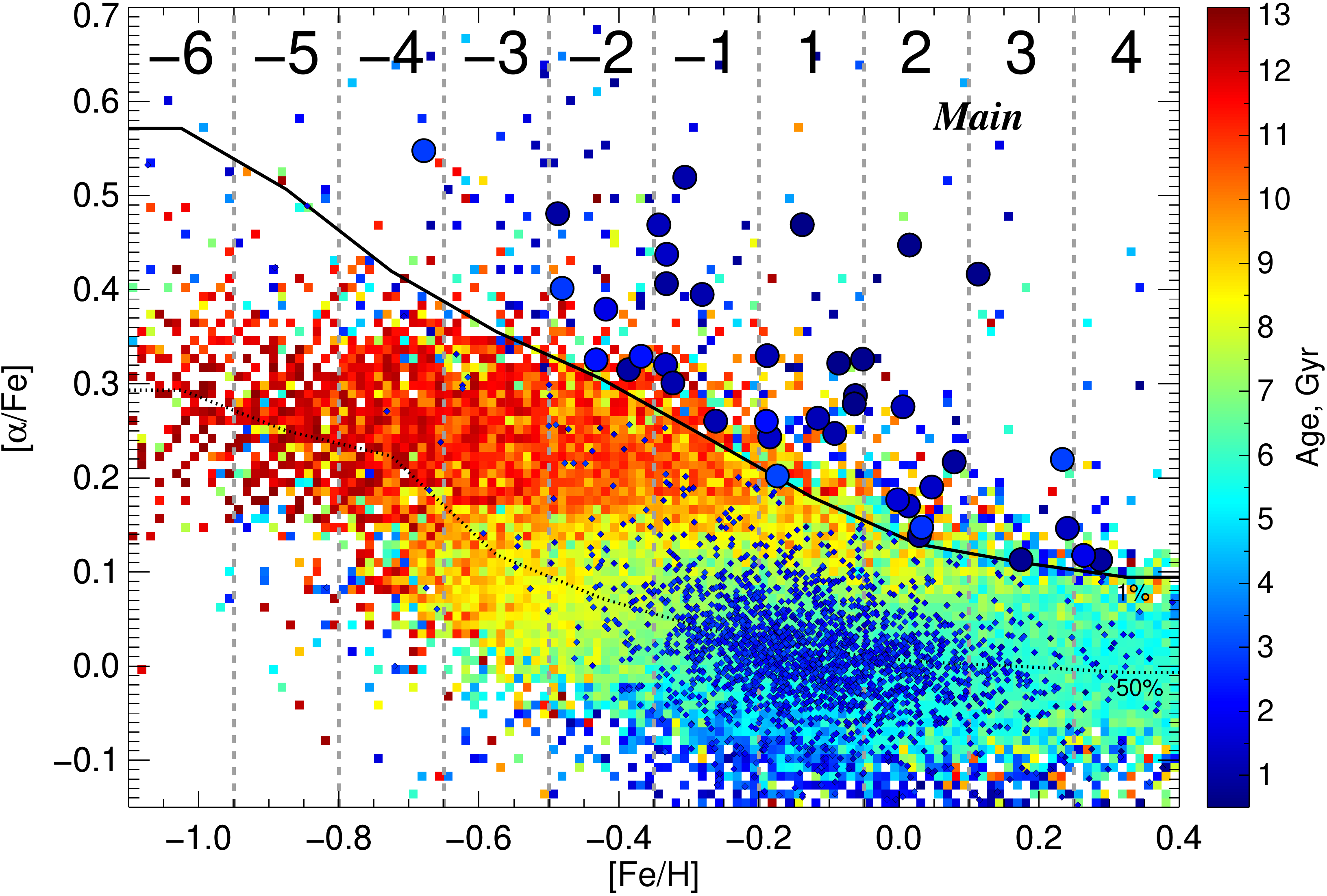}\par
    \includegraphics[width=1\linewidth]{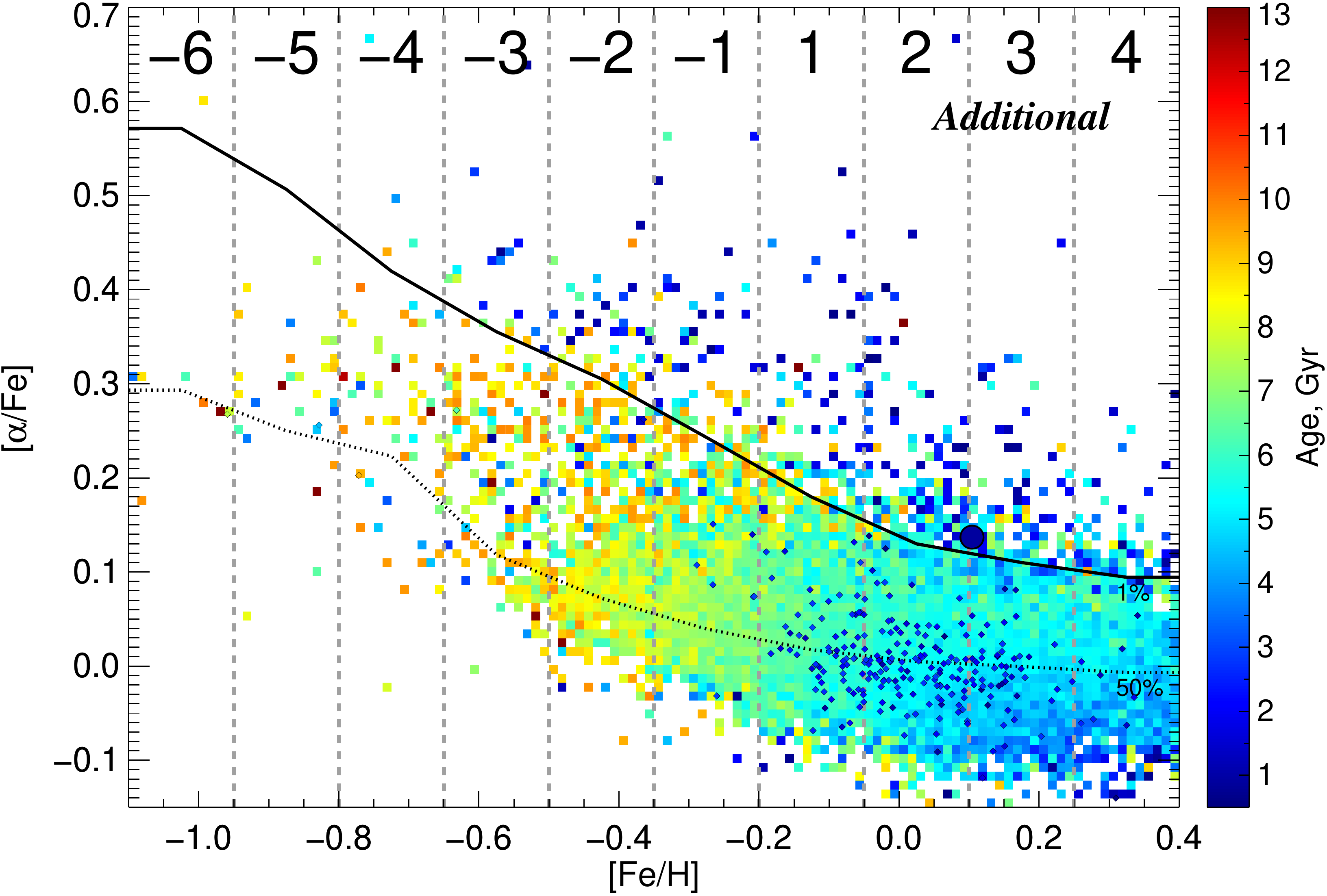}\par
    \end{multicols}
    \begin{multicols}{2}
    \includegraphics[width=1\linewidth]{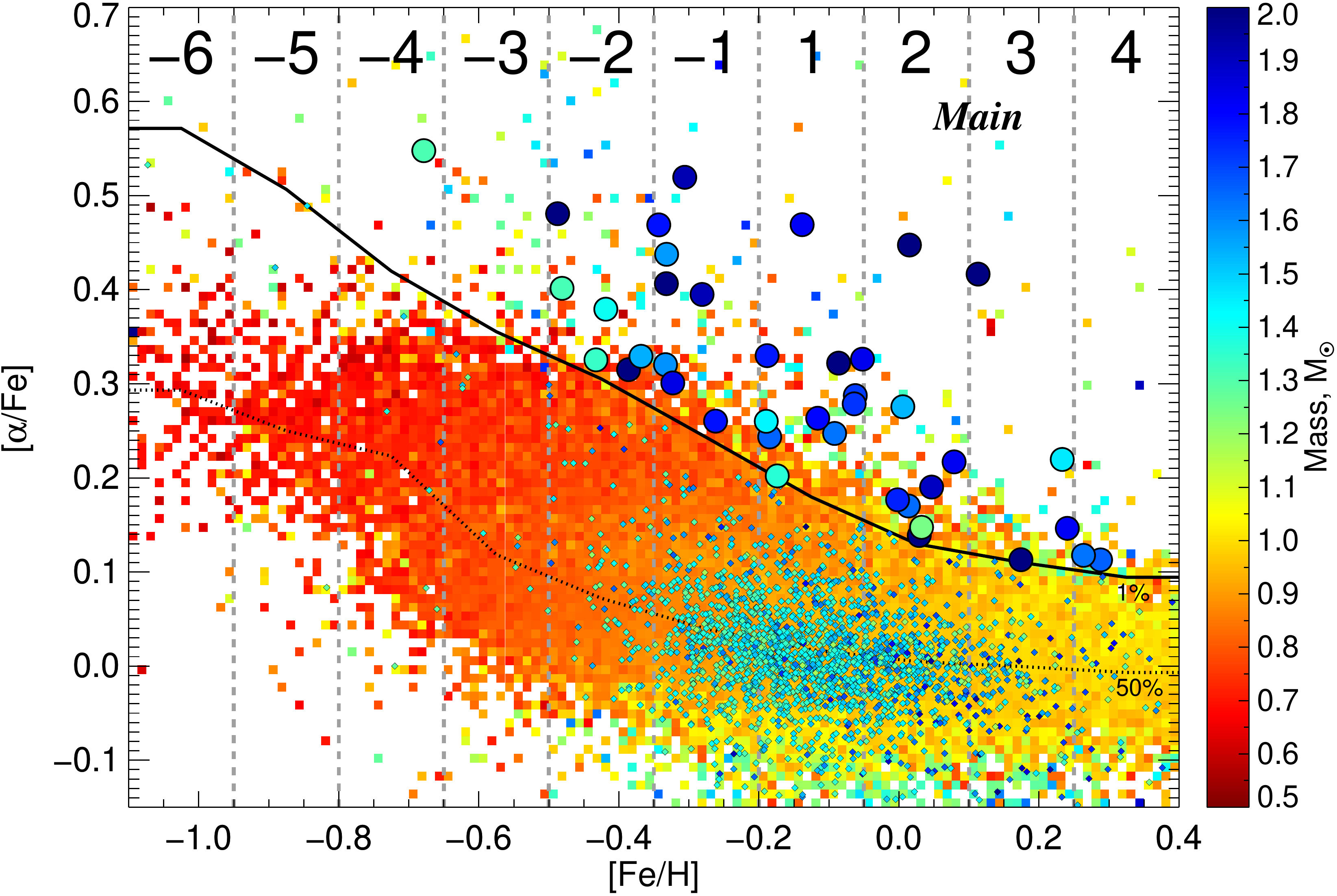}\par
    \includegraphics[width=1\linewidth]{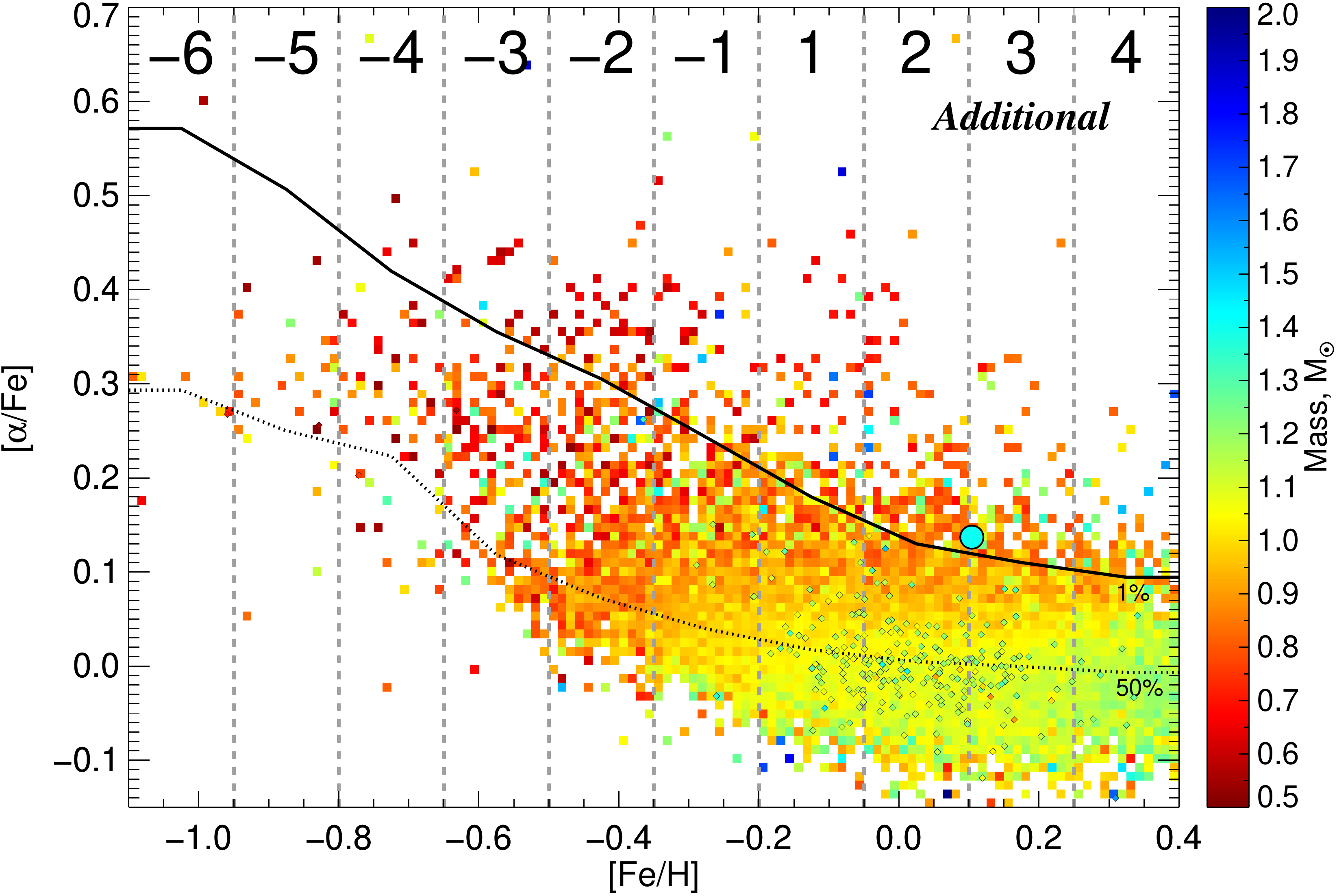}\par
    \end{multicols}
    \begin{multicols}{2}
    \includegraphics[width=1\linewidth]{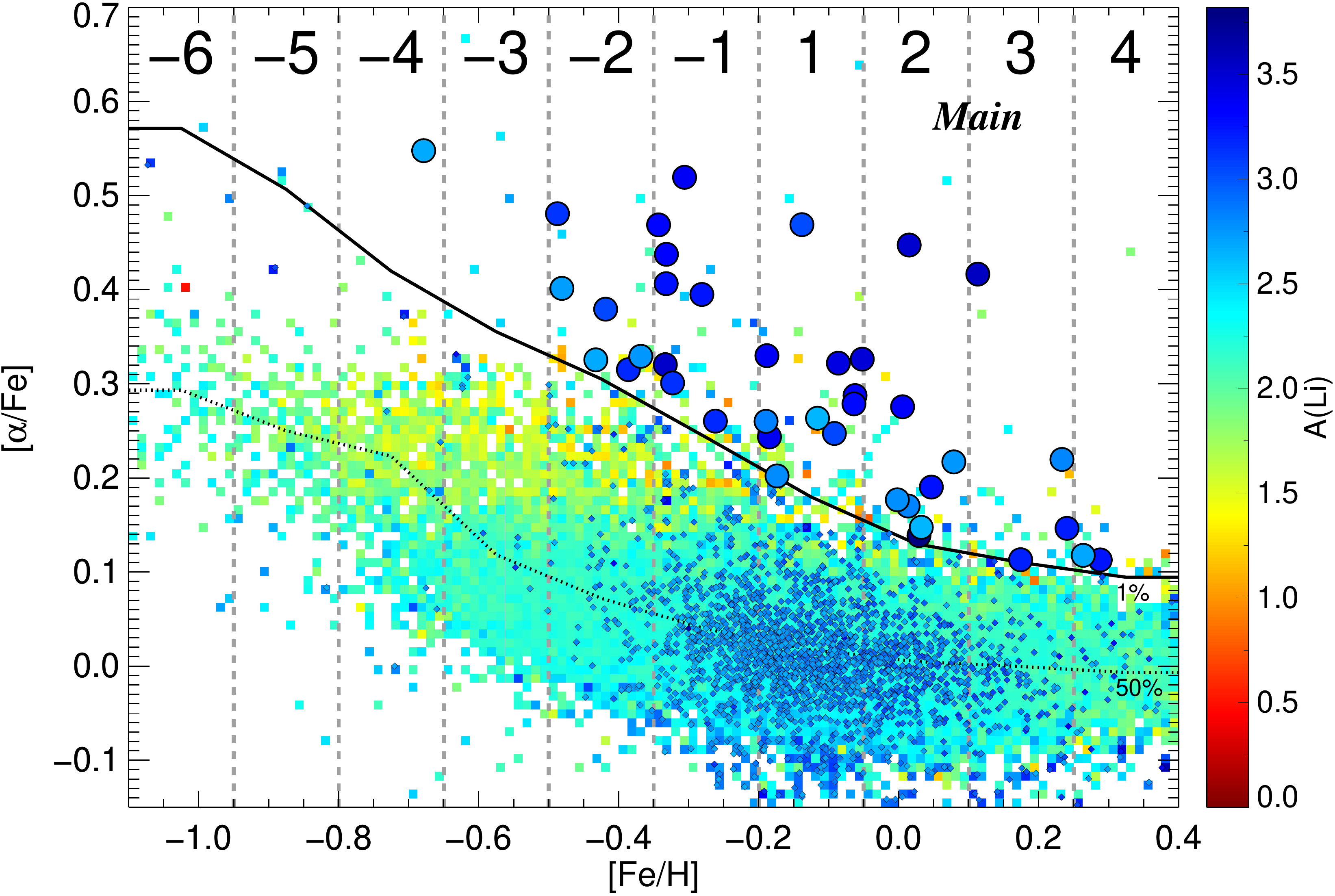}\par
    \includegraphics[width=1\linewidth]{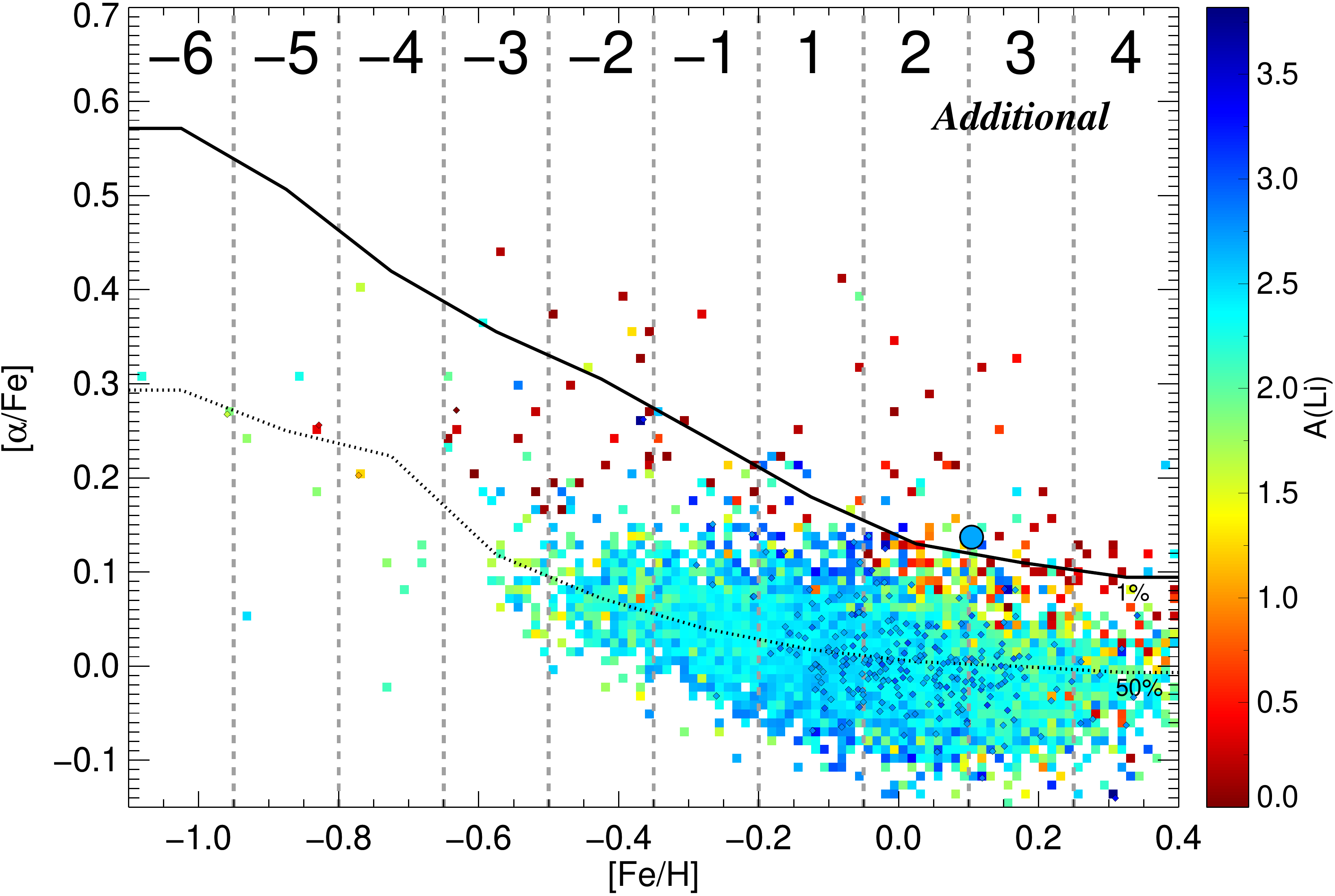}\par
    \end{multicols}
   \caption{Properties of the dwarf stars of the main and additional samples (left and right columns, respectively; see details in Table~\ref{tab_sample}) in the Tinsley-Wallerstein diagram. The solid curve separates the top 1\% of stars in each of the 10 metallicity bins from the remaining 99\% (\exafe and \anormal \ stars, respectively), whereas the dashed curve is the median (50\% of stars above and below it). From top to bottom: Ages, masses, and Li abundances are color-coded, with the color of the background indicating the mean value of the property in bins of size 0.0125~dex in \feh and 0.0094~dex in [$\alpha$/Fe]. Individual stars with age $\leq$~3~Gyr and A(Li)$\geq$~2.65~dex are also indicated by big and small points for stars above and below the top 1\% curve, respectively.}
     \label{afe_vs_feh_age_mass_Li_main_add}
 \end{figure*}

\subsection{Ex$\alpha$fe dwarfs and their properties}
\label{subsect-exafe}

\subsubsection{Defining \exafe dwarfs as outliers in the [$\alpha$/Fe]-[Fe/H] plane}
\label{definitionalpharich}

The positions of the dwarf stars in the \afe vs [Fe/H] diagram are shown in Fig.~\ref{afe_vs_feh_age_mass_Li_main_add} for both the main and additional samples. Here, we follow \citet{Buder2021}, who refer to it as the Tinsley-Wallerstein diagram.

We selected extreme [$\alpha$/Fe]-rich ({{ex$\alpha$fe}}) stars as outliers with the highest \afe values in the main and additional samples described in the previous section. We split the considered \feh range in ten bins of 0.15~dex width, 
keeping bins 1 to 4 at the same positions as in \citet{Charbonnel2021}. In each metallicity bin, we compute the \afe value above which lie 1\% of the stars of the main sample of this bin, namely, stars above the solid curve in Fig.~\ref{afe_vs_feh_age_mass_Li_main_add} (we used the same curve to delineate \exafe both in case of the main and additional samples). In this way, we were able to keep the most ``extreme'' \afe stars of both samples, at more than 2$\sigma$ ($>$0.1~dex) from the median \afe value in each bin, which is indicated by the lower (dashed) curve (50\%) in Fig.~\ref{afe_vs_feh_age_mass_Li_main_add}. The 1\% curve can be approximated by \afe =-0.42$\cdot$\feh+0.13 for \feh<0 and -0.075$\cdot$\feh+0.13 for \feh >0. There are 1~122 and 632 stars from (respectively) the main and additional samples in that region of the \afe vs \feh plane (see columns 3 and 6 -- \exafe and add-\exafe -- in Table~\ref{tab_sample}). In addition to their extreme \afe values, the properties of these stars differ considerably from those of the thick disk, as we argue in the following sections.

Since observational and analysis systematics could lead to a degeneracy between the determination of [Fe/H] and that of [$\alpha$/Fe], we conducted a test to check if this could shift \anormal \ stars to the \exafe region in the Tinsley-Wallerstein diagram. We selected 584 GALAH~DR3 stars from nine open clusters of various ages and metallicities using the same selection criteria as described in \S\ref{stellar_parameters_abundances}. 
For each cluster, we computed the mean \feh and [$\alpha$/Fe] as well as the \feh and [$\alpha$/Fe] deviation of individual member stars which should in principle share the same initial composition. Combining deviations of individuals of all the clusters, we derived the degeneracy trend in the Tinsley-Wallerstein diagram and found a slope of -0.36$\cdot$[Fe/H], which is lower than the approximate slope of the 1\% curve of (-0.42$\cdot$[Fe/H]) mentioned in the previous paragraph. Hence, the degeneracy can possibly lead to a slight increase in \afe if \feh is underestimated. However, it  cannot ``shift'' \anormal \ stars to the \exafe region above the 1\% curve, unless the stars are already close to this limit.

\subsubsection{Ages, masses, and Li content of \exafe dwarfs}
\label{agealpharich}

In Fig.~\ref{afe_vs_feh_age_mass_Li_main_add}, we display some properties of the {main} and {additional} sample stars in Tinsley-Wallerstein diagram. In the top panels, we indicate the stellar ages, color-coded. We adopted bins of 0.0125~dex in \feh and 0.0094~dex in \afe and indicate the mean age of each bin. As expected, both on theoretical and observational grounds, the \afe values decrease with decreasing age of stars and increasing metallicity, ranging from more than $\sim$10~Gyr on the top left (thick disk) to less than a few Gyr in the bottom right (thin disk). \afe is a good proxy for age, at least for sub-solar \feh values. However, our \exafe sample defined in the previous section appears clearly to go against that trend, being dominated by young ages -- and clearly younger than the bulk of the stars in most metallicity bins. 
 
\begin{figure}
     \includegraphics[width=\linewidth]{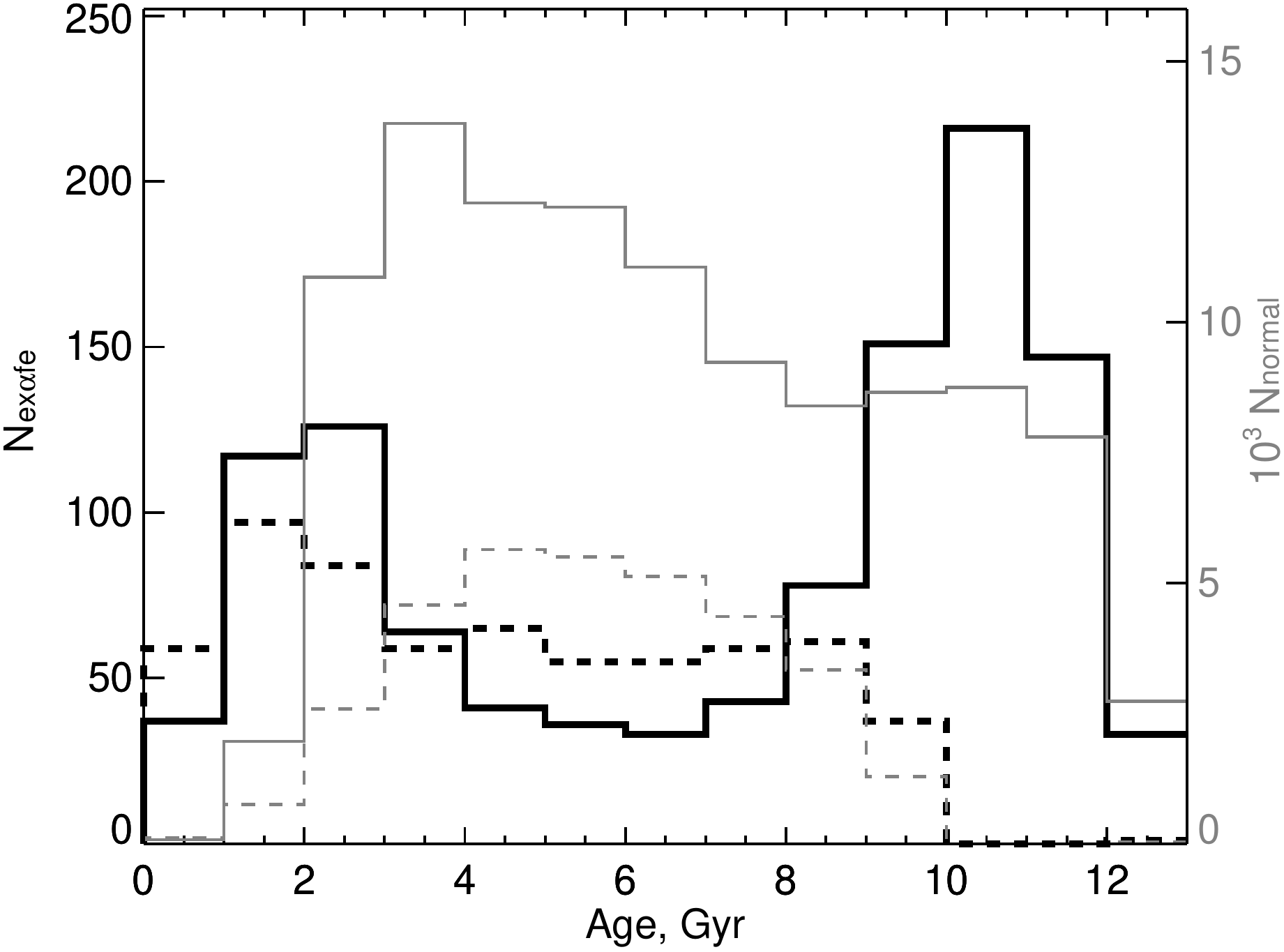}
     \includegraphics[width=\linewidth]{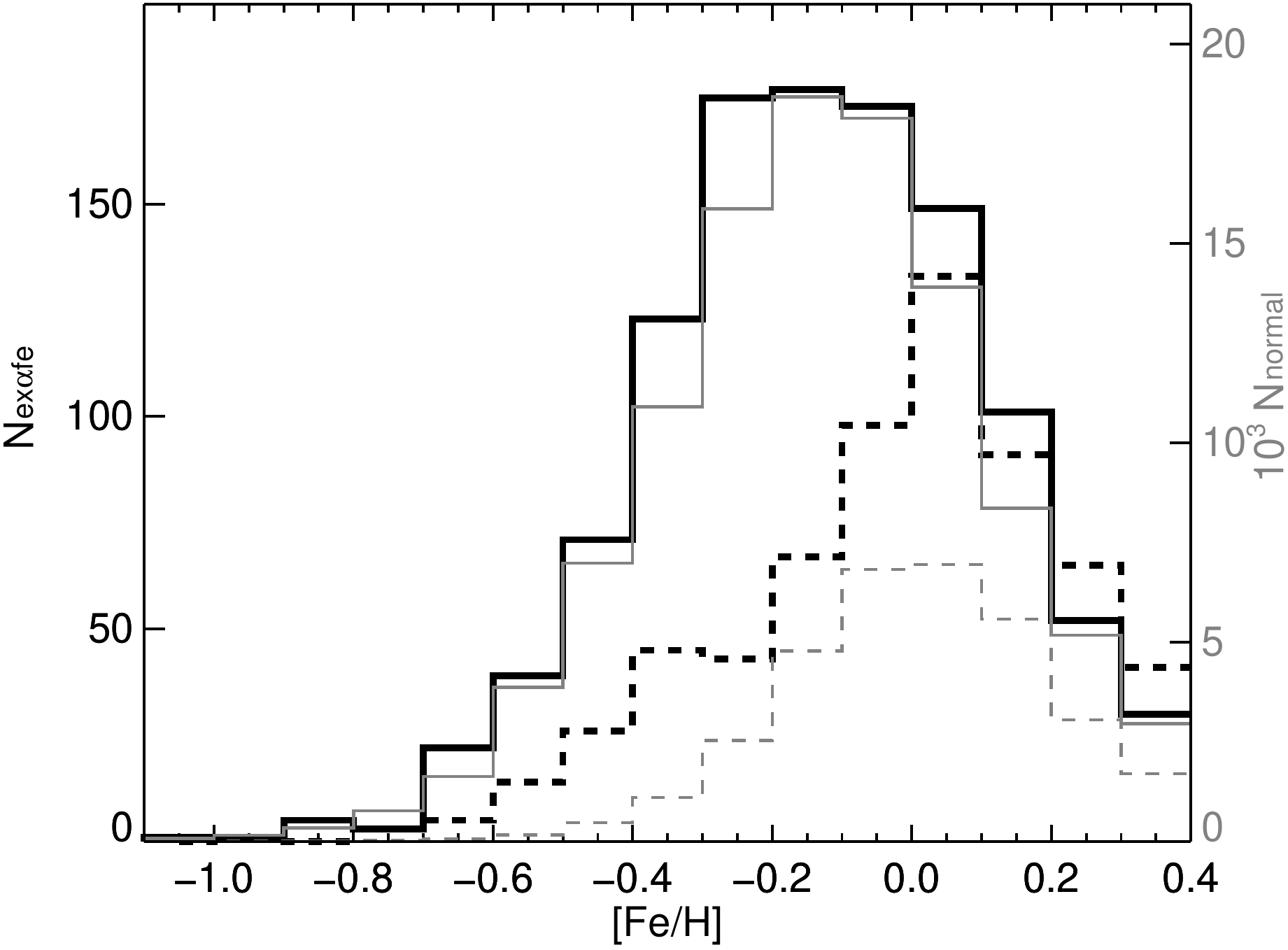}
     \caption{Age ({\it top}) and [Fe/H] ({\it bottom}) distributions of the \exafe (thick black, scale on the left) and \anormal \ (thin grey, scale on the right) dwarf stars from the main (solid) and additional (dashed) samples.}
      \label{hist_age_feh_alpha_rich}
 \end{figure}

Among the 1~122 \exafe stars of the main sample, $\sim$20\% (280) are younger than 3~Gyr and we define them as \exafey stars (Young stars with EXtreme \afe ratio, see the fourth column \exafey of Table~\ref{tab_sample}).
The reasoning behind our choice to consider them to be a class apart can be clearly seen in Fig.~\ref{hist_age_feh_alpha_rich} (top panel). While the \anormal \ sample displays a rather uniform age distribution (thin grey histogram), the \exafe sample (thick black histogram) displays two prominent regions: one at high ages (9-12~Gyr) which can, in principle, be identified with the thick disk; and another one at low ages, which is unexpected in view of their high \afe values. In contrast, the \anormal and \exafe samples are indistinguishable in their metallicity distributions (bottom panel in Fig.~\ref{hist_age_feh_alpha_rich}), both peaking at slightly sub-solar metallicity. These \exafey stars constitute the main topic of our work and we postpone a specific discussion of them in the next section, after surveying the other properties of our samples (mean properties in the aforementioned [$\alpha$/Fe]-[Fe/H] bins).

In the top right panel of Fig.~\ref{afe_vs_feh_age_mass_Li_main_add}, we display the ages of our add-\exafe sample. The trends in age are similar to the case shown in the top left panel for the main sample, although the oldest ages ($>$10~Gyr) appear to be missing. As explained in Sec.~\ref{ages_sample}, this is due to the higher age uncertainties ($\sigma_\mathrm{age}$/age$>$30\%) of that additional sample, which contains mostly low-mass stars lying close to the ZAMS. This is clearly shown in the top panel of Fig.~\ref{hist_age_feh_alpha_rich} (dashed histograms) for both the add-\anormal \ sample and the add-\exafe sample, the latter displaying an enhancement of its young population, as in the case of the main \exafe sample (solid histogram). The metallicity distributions of the add-\anormal \ sample and the add-\exafe sample are similar and peak at solar values (bottom panel of Fig.~\ref{hist_age_feh_alpha_rich}, dashed histograms). The shift in [Fe/H] between the main and additional samples is artificially induced by the lack of older, more metal-poor stars in the second one. When we combine both samples, the [Fe/H] distributions of the \exafe and \anormal \ dwarfs perfectly overlap with a peak at -0.1--0~dex. 

The middle panels of Fig.~\ref{afe_vs_feh_age_mass_Li_main_add} display the masses of the dwarf stars of our samples. The age trend of the top panels is now turned into a mass trend, with the older ages corresponding to lower masses, as expected: lower average masses are found at the lowest metallicities and higher \afe ratios. 
However, in the left middle panel, the \exafe sample (above the solid curve) again shows  considerably higher average masses than the stars with lower \afe values immediately below the solid curve, in agreement with the young ages of the main \exafe stars in the top panel. Those stars are missing from the add-\exafe sample and, thus, the stars above the solid curve in the right middle panel are mostly of a low mass. There is a clear consistency between the picture emerging from the top and middle panels of Fig.~\ref{afe_vs_feh_age_mass_Li_main_add}.

In the bottom panels of Fig.~\ref{afe_vs_feh_age_mass_Li_main_add}, we display another key property of the stars of our samples, color-coding them according to their Li abundance. The interpretation of those panels is more difficult since their surface Li content depends on a combination of the two previous properties -- age and mass -- as well as their initial Li content, which is unknown because it depends on the chemical evolution of Li: in principle, the initial value of every star should be above the primordial one of A(Li)$_P$=2.65 \citep{Pitrou2018}. Additionally, substantial Li depletion may occur along the main sequence and even the pre-main sequence for low-mass stars \citep[e.g.][]{Magazzu1992,Deliyannisetal2000IAUS,Castroetal2016,Dumontetal2021b,Jeffries2021}. 
The bulk of our sample stars has mean values A(Li)$\sim$2.5~dex, lower than A(Li)$_P$, in all the bins of the \afe vs \feh diagram. For \afe$>$0.2 and \feh$<$-0.4 (realm of thick disk) they become even lower, less than 1.5~dex.

The \exafe stars of the main sample (bottom left panel of Fig.~\ref{afe_vs_feh_age_mass_Li_main_add}) display higher mean values of Li than the stars of the thick disk. This is consistent with their higher average mass and lower average age (hence lower Li depletion), already discussed in the previous paragraphs. On the other hand, the \exafe stars in the add-sample (bottom right panel) display lower average Li values than the other stars of that sample, because they correspond to lower average masses, as shown in the middle right panel and discussed above. In fact, the stars of the thick disk in that panel (corresponding to \feh$<$-0.1 and \afe$>$0.2) have no detectable Li values and are absent from this region of the diagram.

In summary, the \exafe stars of our main sample are, on average, younger and more massive than the remaining ones (i.e., the non-ex$\alpha$fe), and they are more (or at least as much) Li-rich. The \exafe stars of the additional sample are also younger on average, but less massive;  for that reason, they are also Li-poorer than the remaining ones (the Li behavior in the \exafe stars will be discussed further in \S~\ref{high_li}). As explained above, the differences in terms of stellar masses between the main and additional \exafe samples come directly from the isochrone age-dating method which precision depends on the positions of the stars in the HRD. The identification of these biases is of importance for compiling a relevant description and understanding of the \exafe population.

\subsection{Young-\exafe dwarfs}
\label{propertiesyoungalpharich}
In the previous section, we have identified a minority of stars -- about one percent in the main and two percent in the additional samples of dwarf stars -- with unexpectedly high values of \afe for their metallicity. We have shown that some of those stars -- about 25\% for the main and 40\% for the additional sample -- are also young, younger than 3~Gyr (\exafey stars). The existence of a population of ``young'' red giant stars with ``high [$\alpha$/Fe]'' has already been reported with APOGEE data \citep[e.g.,][]{Chiappini2015,Martig2015,Anders2017,2018MNRAS.475.5487S} and LAMOST \citep{Zhang2021}, albeit with very different limits regarding \afe and age than adopted here (\S~\ref{high_alpha_giants}). It has been argued that the properties of those stars should be interpreted by considering that they are, in fact, old stars with high \afe which have merged recently: this would explain their higher than average mass reported by previous studies, and make them appear younger (e.g., \citealt{Zhang2021,Miglio2020}). 

In the following, we further explore  the various properties of the \exafey dwarf population, namely, mass, Li abundance and kinematics, and we show that the interpretation in terms of mergers cannot hold. We also discuss the case of \exafey giants in \S~\ref{high_alpha_giants}.

\subsubsection{Ages, metallicities, and masses}
\label{subsec:ages_met_masses}
We overplot in each panel of Fig.~\ref{afe_vs_feh_age_mass_Li_main_add} the \exafe stars with young ages ($<$3~Gyr) and Li abundances higher than the primordial BBN value of A(Li)$_P$=2.65 \citep{Pitrou2018}. They are displayed as individual large dots (and not as average values for each pixel, as was done for the other stars) in the top part of each panel (the \exafe region above the solid curve). For the purposes of comparison, we also plotted as small dots the individual values for all the stars younger than 3~Gyr of our main and additional samples in all the panels. For all stars and panels, the color-coding is the same as adopted in the previous section and indicated on the vertical bars on the right.

In all the panels, the large majority of young stars are positioned where they are expected to be, namely, the thin disk, around -0.3$<$\feh$<$0.2~dex and -0.1$<$\afe$<$0.1~dex (small dots). These stars have also high Li content, as expected, and higher mass than the average of their pixel. Among the \exafe we also find similar stars which at the same time have young age, high \afe and high Li (large dots). 

\begin{figure}[h]
     \begin{multicols}{2}
      \includegraphics[width=\linewidth]{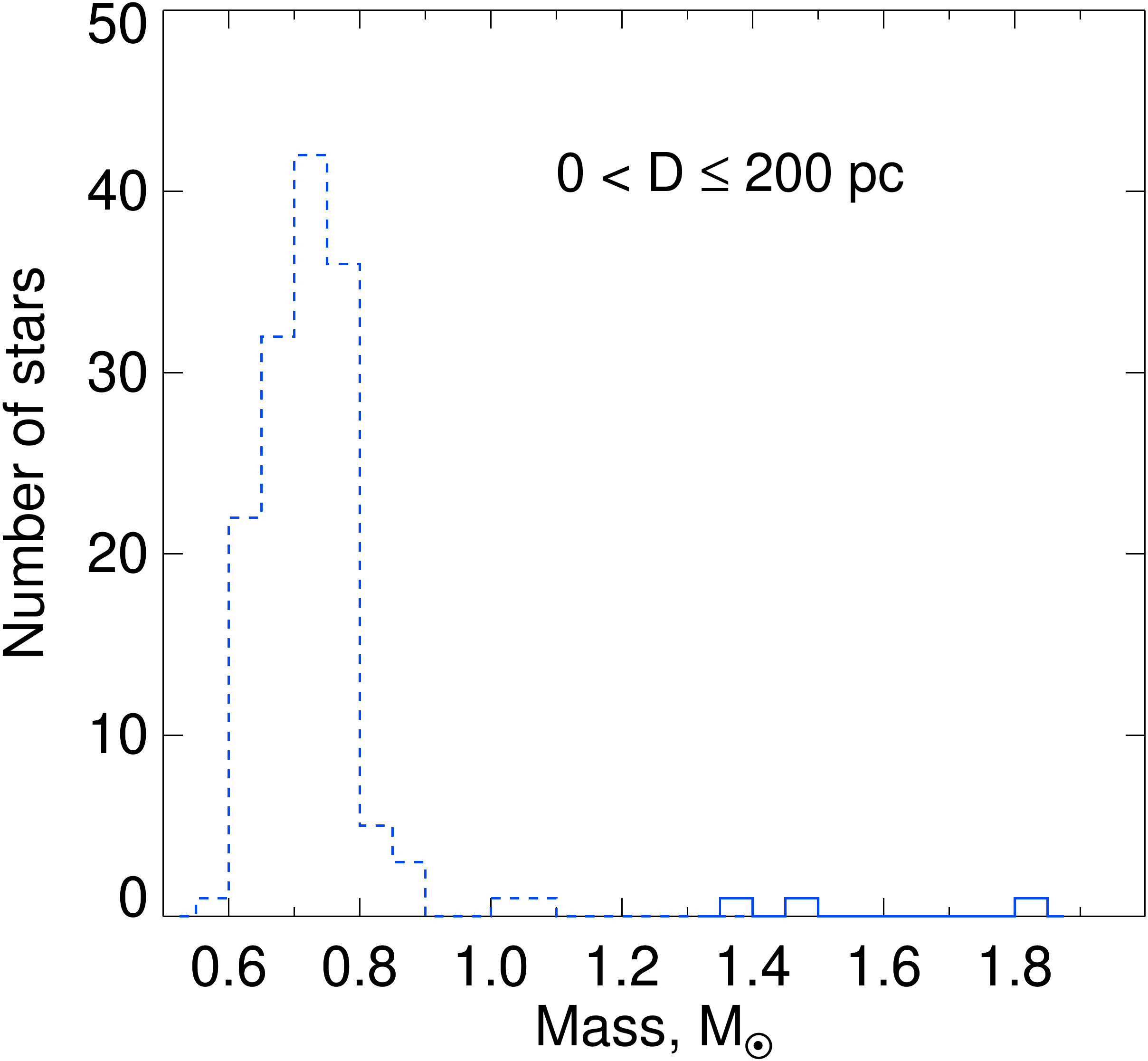}\par    
     \includegraphics[width=\linewidth]{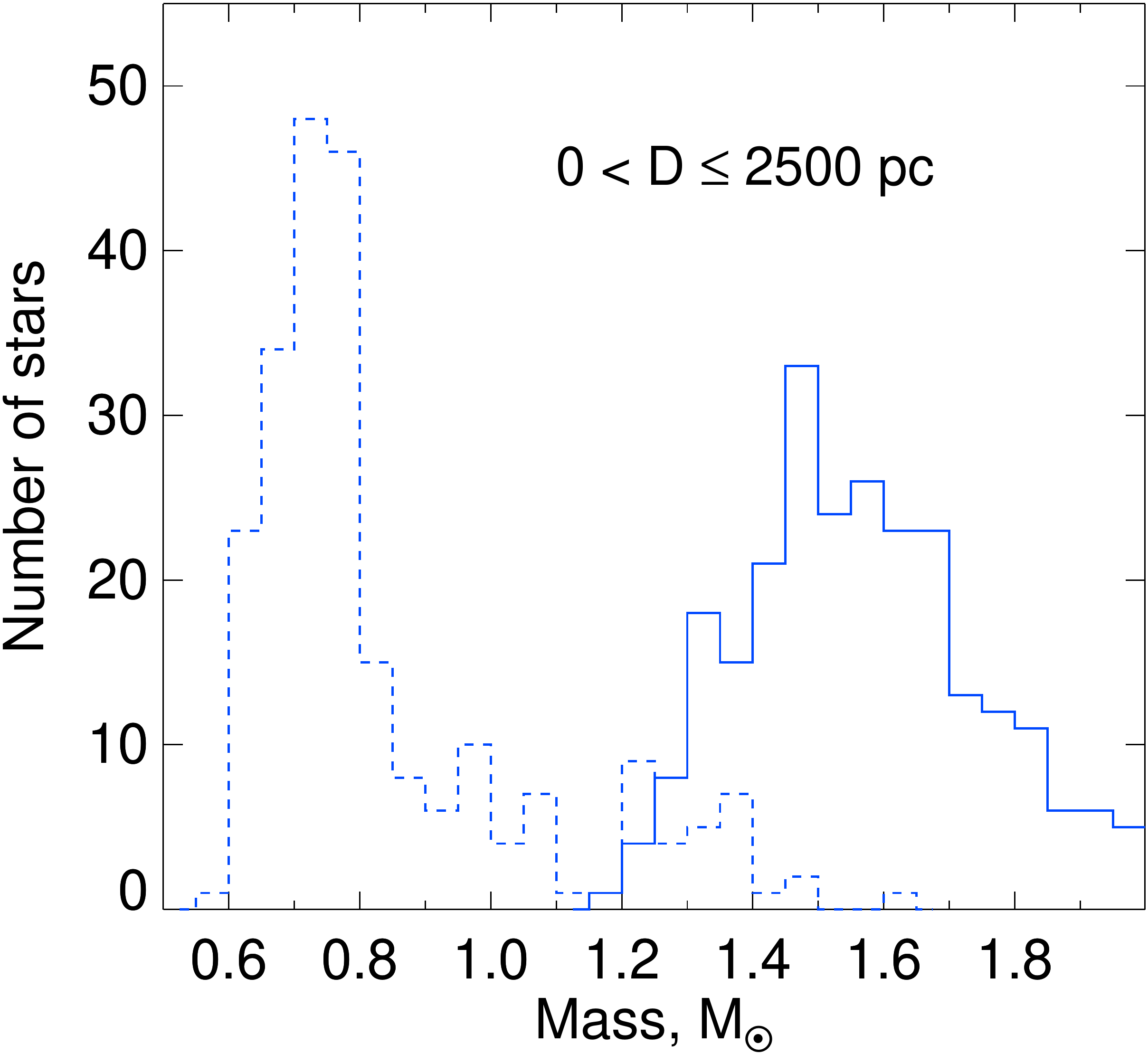}\par    
     \end{multicols}
     \begin{multicols}{2}
     \includegraphics[width=\linewidth]{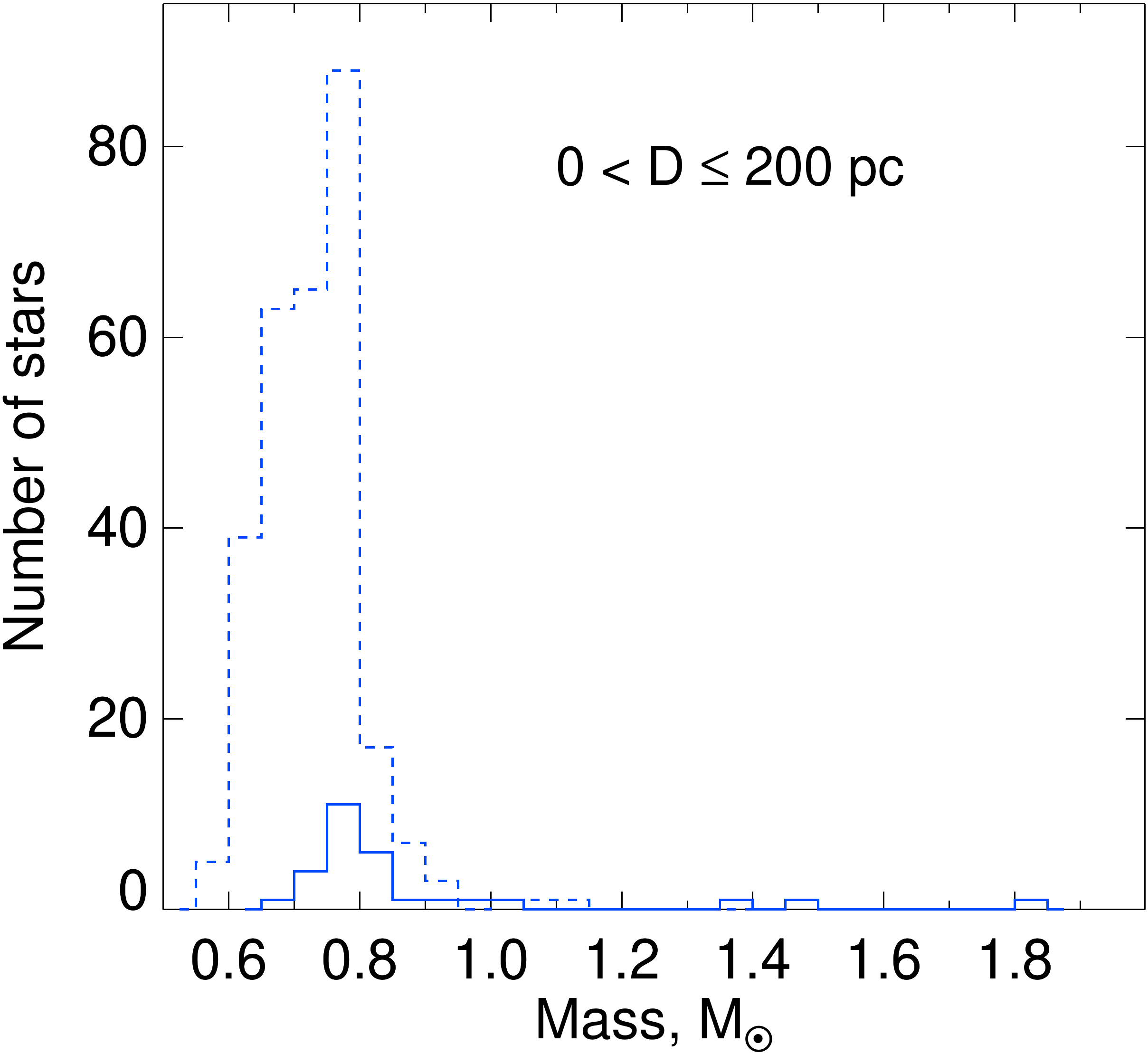}\par        \includegraphics[width=\linewidth]{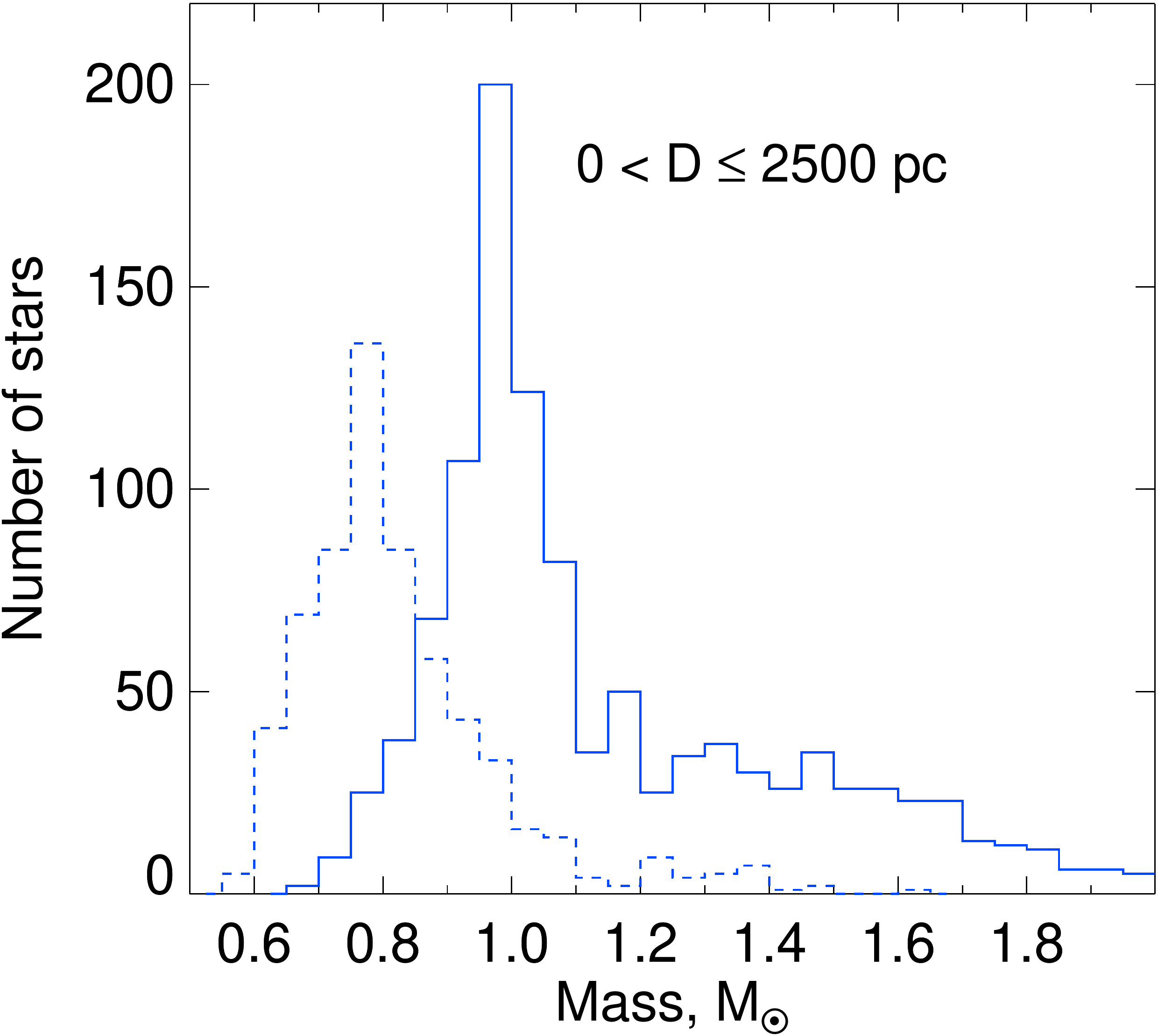}\par
    \end{multicols}  
     \begin{multicols}{2}
     \includegraphics[width=\linewidth]{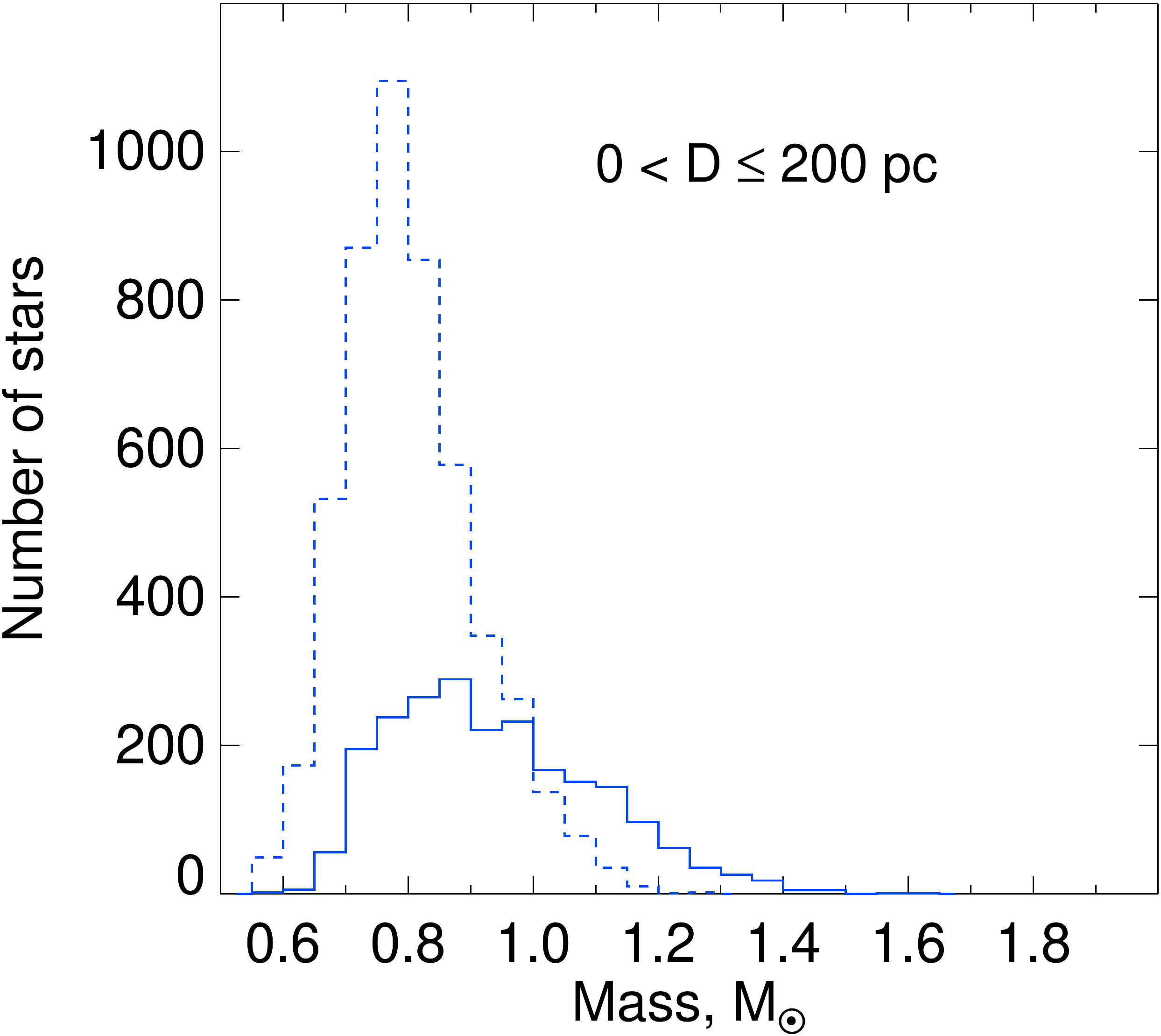}\par  
         \includegraphics[width=\linewidth]{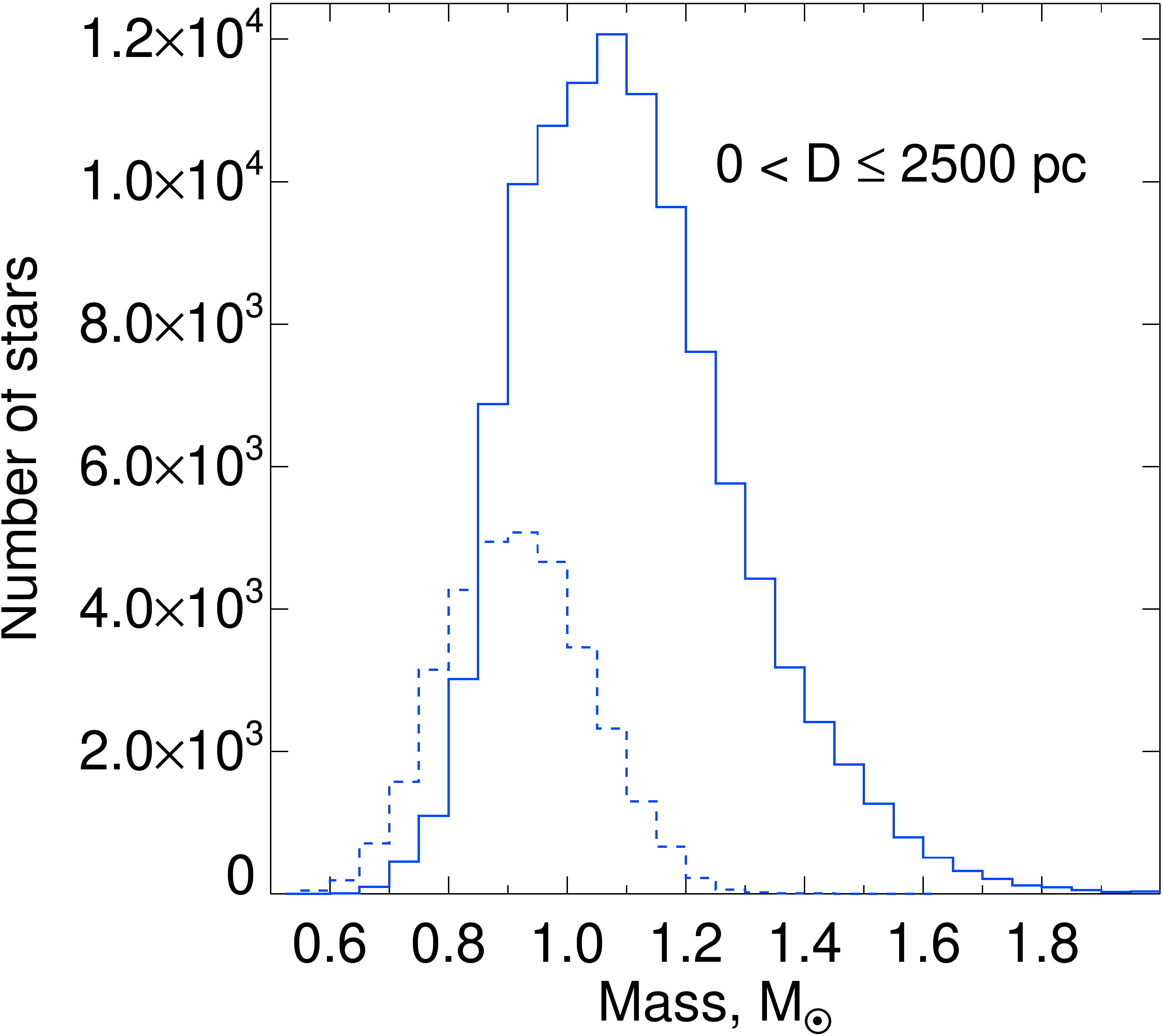}\par 
      \end{multicols}  
    \caption{Mass distribution of \exafey (top), all \exafe (middle) and \anormal \ (bottom) stars. Nearby stars (distance D$<$200 pc) are on the left panels and all stars up to 2500 pc on the right panels,  for the main (solid) and additional (dotted) samples, respectively.}     
\label{fig:mass_distr_all}     
 \end{figure}

The mass distribution of the \exafey dwarf stars appears in the top panels of Fig.~\ref{fig:mass_distr_all}, both for those of the main sample (solid histograms) and of the additional sample (dotted). To interpret these figures, it is important to keep in mind that the GALAH survey is magnitude-limited (see \S~2.1 and Fig.~4a in \citealt{Buder2021}): on the one hand, it contains almost no stars brighter than $V=9^m$, implying that massive (luminous) dwarfs stars are absent at small distances. On the other hand, the sample has also a lower luminosity limit -- the $V$-band distribution drops very steeply at $V=14^m$ -- and it contains only $\sim$1\% of dwarf stars that are fainter. As a result, the relative number of low-mass dwarfs (both \exafe and \anormal) decreases at larger distances. Indeed, the additional \exafey sample is composed of small mass stars ($<$1~\ms), which dominate the volume close to the Sun (distance d$<$200~pc, top left) because of their large number in the initial mass function. The main \exafey sample is composed of rather massive stars (M$>$1.2~\ms), which are less numerous because of the IMF; however, they are visible at larger distances and dominate completely in the distance range 500$<$d (pc)$<$2500. As a result of the selection criteria we used, as well as the characteristics of the GALAH survey, the total sample of \exafey stars (main plus additional for d$<$2500 pc) shows a bimodal behavior, as seen in the top right panel. We checked that this bimodality also appears for the \exafe sample, albeit to a smaller degree (middle panels in Fig.~\ref{fig:mass_distr_all}), and with a shift of the high mass peak due to age limit considered as more massive stars have shorter MS lifetimes. The bimodality persists at still lower level for the \anormal \ stars (bottom panels). It is thus an artificial pattern that results from the adopted selection criteria applied to a magnitude-limited survey.

 The conclusion of this analysis asserts that: a) the dwarf \exafey sample contains more massive and thus more luminous (and therefore seen to larger distances) than the older \exafe sample due to secular evolution effects (3~Gyr is approximately the main sequence lifetime of $\sim$1.2-1.6~M$_{\odot}$ stars depending on their metallicity); and b) once the selection biases discussed above are accounted for, \exafey dwarfs do not appear to have higher than average mass compare to their \exafe and \anormal \ counterparts.

\subsubsection{Importance of Li for \exafe dwarfs}
\label{high_li}

\begin{figure}[h]
    \center
    \includegraphics[width=1\linewidth]{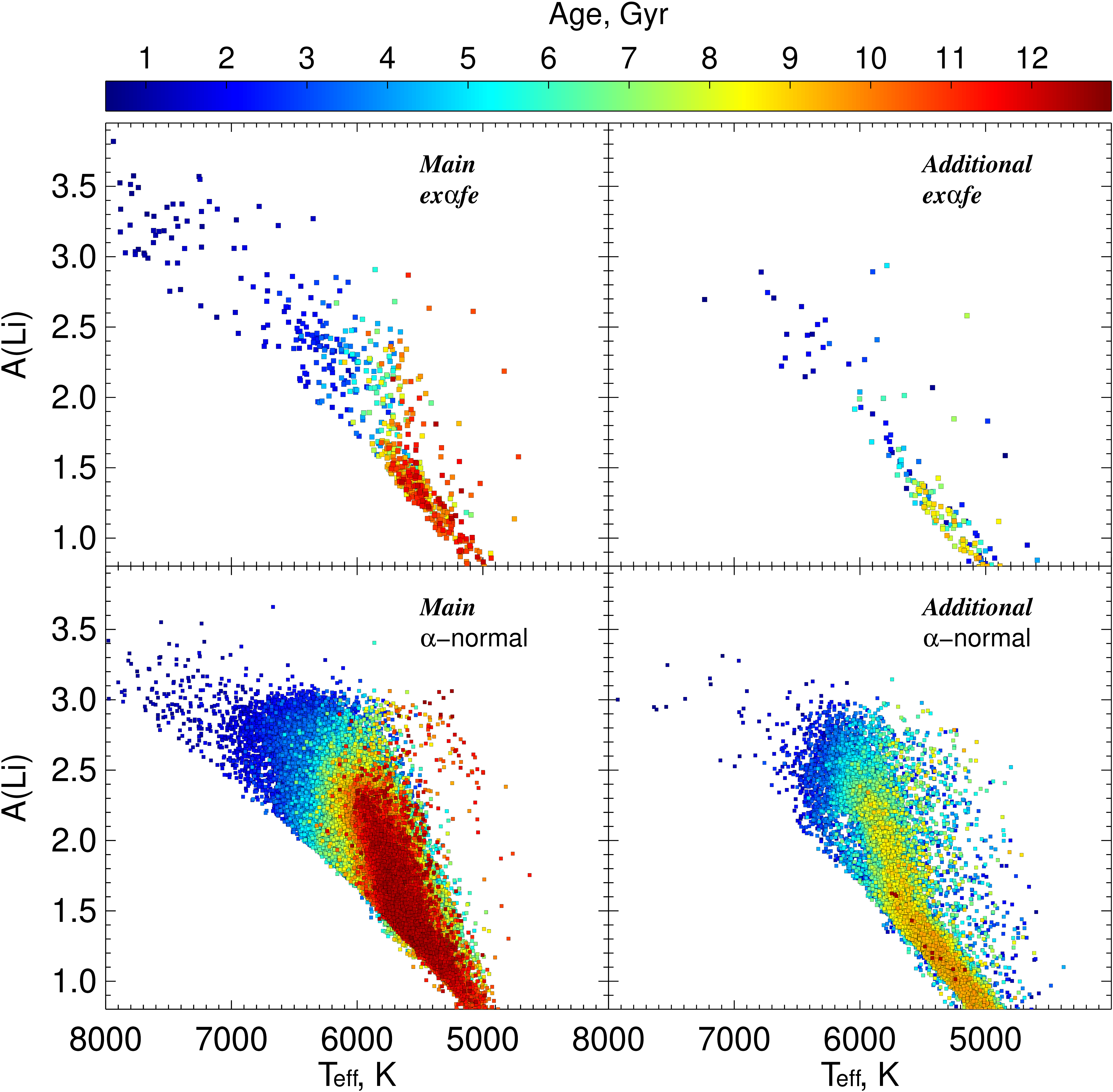}
    \caption{Age color-coded A(Li) values as a function of $T_\mathrm{eff}$ for {{ex$\alpha$fe~}} and \anormal \ stars (upper and lower panels, respectively). Left and right columns are for the main and additional samples, respectively.}
    \label{li_vs_teff_main_merged}
\end{figure}

In Fig.~\ref{li_vs_teff_main_merged}, we display the Li abundance (when available) vs $T_\mathrm{eff}$ of the \exafe and \anormal \ dwarf stars (top and bottom panels, respectively) of the main and additional samples (left and right panels, respectively). The GALAH data are color-coded as a function of age. We note the similarities between the \exafe and the \anormal \ stars. First, the highest Li abundance is similar in both cases, with A(Li)$\sim$3.5, and it is found in the hottest (i.e., more massive) and youngest (\exafey and \anormal) stars. Second, the Li abundance decreases with increasing stellar age and decreasing effective temperature, with the latter being a proxy for the mass. This well-known behavior has long been observed in field and open cluster dwarfs \citep[e.g.,][]{zappala1972,sestito2005}. It is interpreted as the result of internal transport processes of chemicals which lead to differential photospheric Li depletion in stars of different masses and ages along the main sequence (and eventually already on the pre-main sequence for the lowest masses). Notably, Li depletion is minimal or eventually null in early-F and late A-type stars that exhibit the highest Li abundances close to the value they were born with \citep[][and references therein]{Charbonnel2021}.

We conclude from this analysis that: 1) the Li-abundance of the hottest \exafey dwarf stars is slightly (by a factor of 2) above the proto-solar value of A(Li)=3.26~dex and constitutes a strong argument for a young age; 2) the ages of the stars with the highest Li abundances, independently determined, are indeed low (less than a few Gyr); 3) Li depletion occurs similarly in \exafe and \anormal \ dwarfs; 4) the \exafe and \anormal \ populations were born with essentially the same maximum initial Li abundances. 

The high Li content of the \exafey stars, which is similar to that of the young \anormal \ stars, makes it difficult to adopt the merger scenario as an explanation for their young age. The merger would indeed have to restore the exact original Li content of the two components, which would certainly require strong fine-tuning. In our opinion, this is a decisive argument against that scenario; however, it is not the only one.
It is clear that old stars with high effective temperatures and high values of A(Li) are not generally observed, while the $\alpha$-Li-richest stars (squares in the figure) also follow the general trend.

\begin{figure}[h]
     \includegraphics[width=\linewidth]{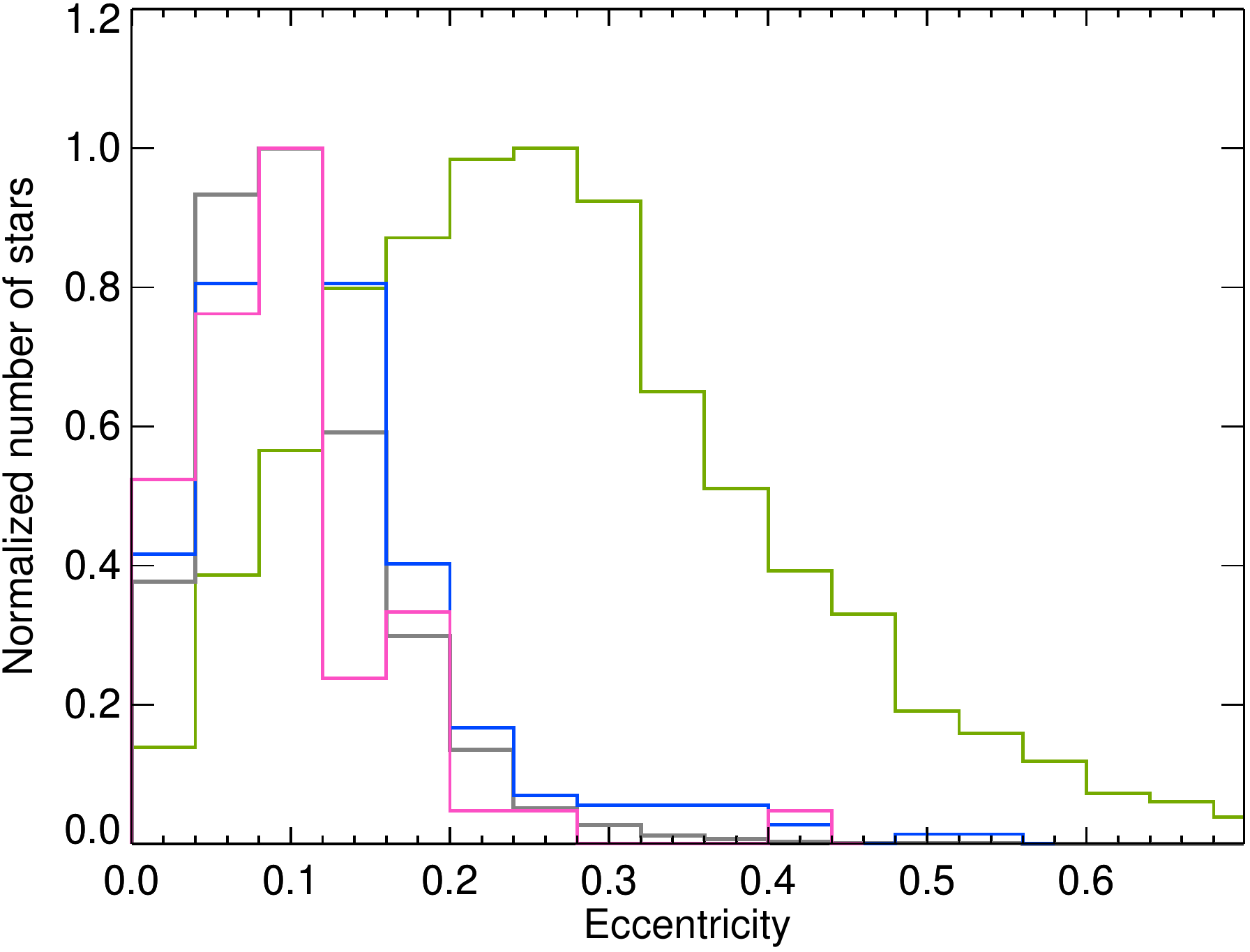}
     \includegraphics[width=\linewidth]{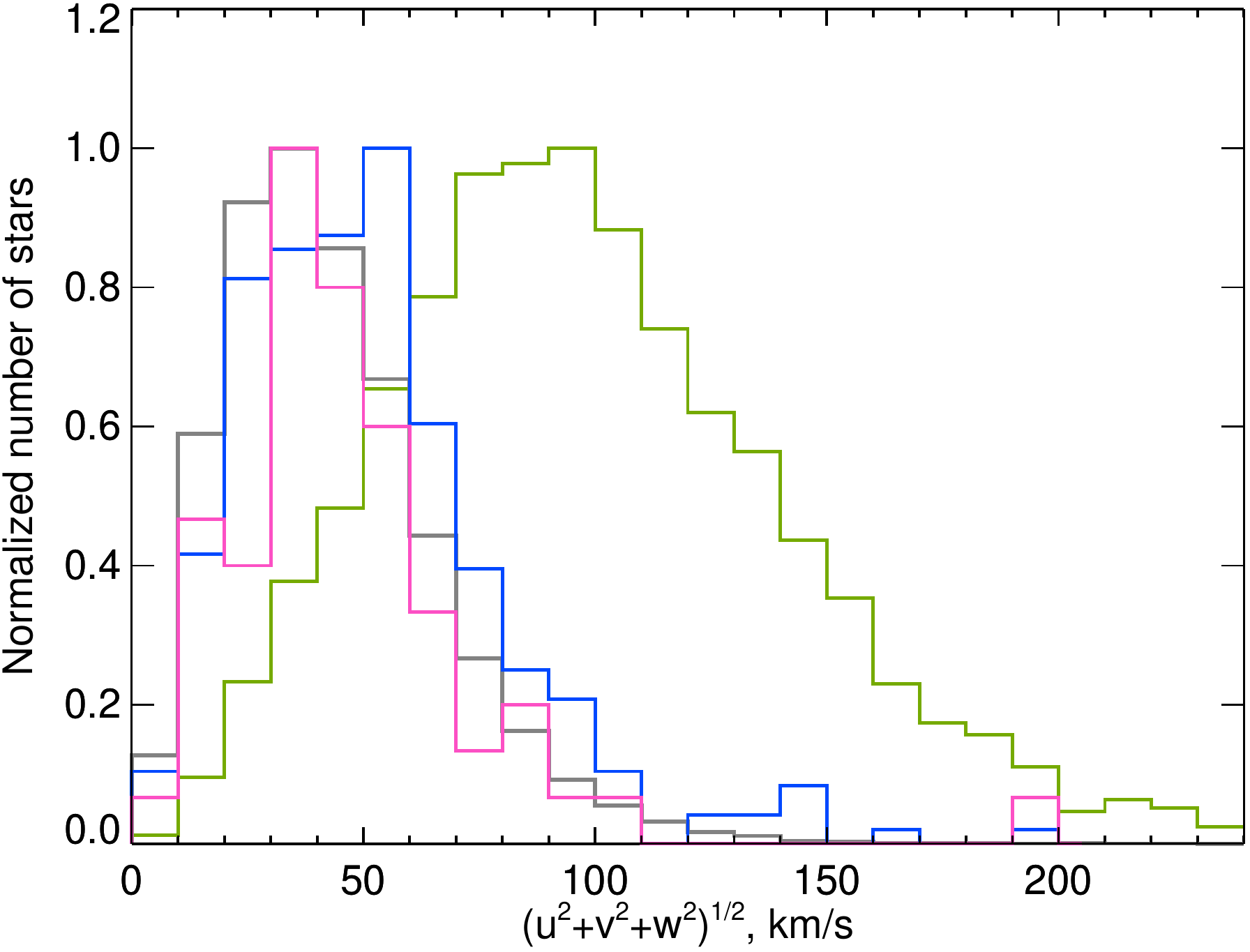}
     \caption{Distribution of eccentricities and velocities with respect to the local standard of rest $V_{LSR}$ (upper and bottom panels, respectively) for \exafey and young \anormal \ dwarfs stars (blue and grey histograms, respectively). The green histogram shows the distribution of old (age$\geq$8~Gyr) stars with \afe$\geq$0.2~dex. The magenta histogram shows the distribution of 40 \exafey stars with A(Li)$>$A(Li)$_{SBBN}$=2.65. All the histograms are scaled to have the same height and we show the distributions for the main sample only for the purpose of better visibility.}
    \label{vel_ecc_distr}
\end{figure}

\begin{figure}[h]
     \includegraphics[width=\linewidth]{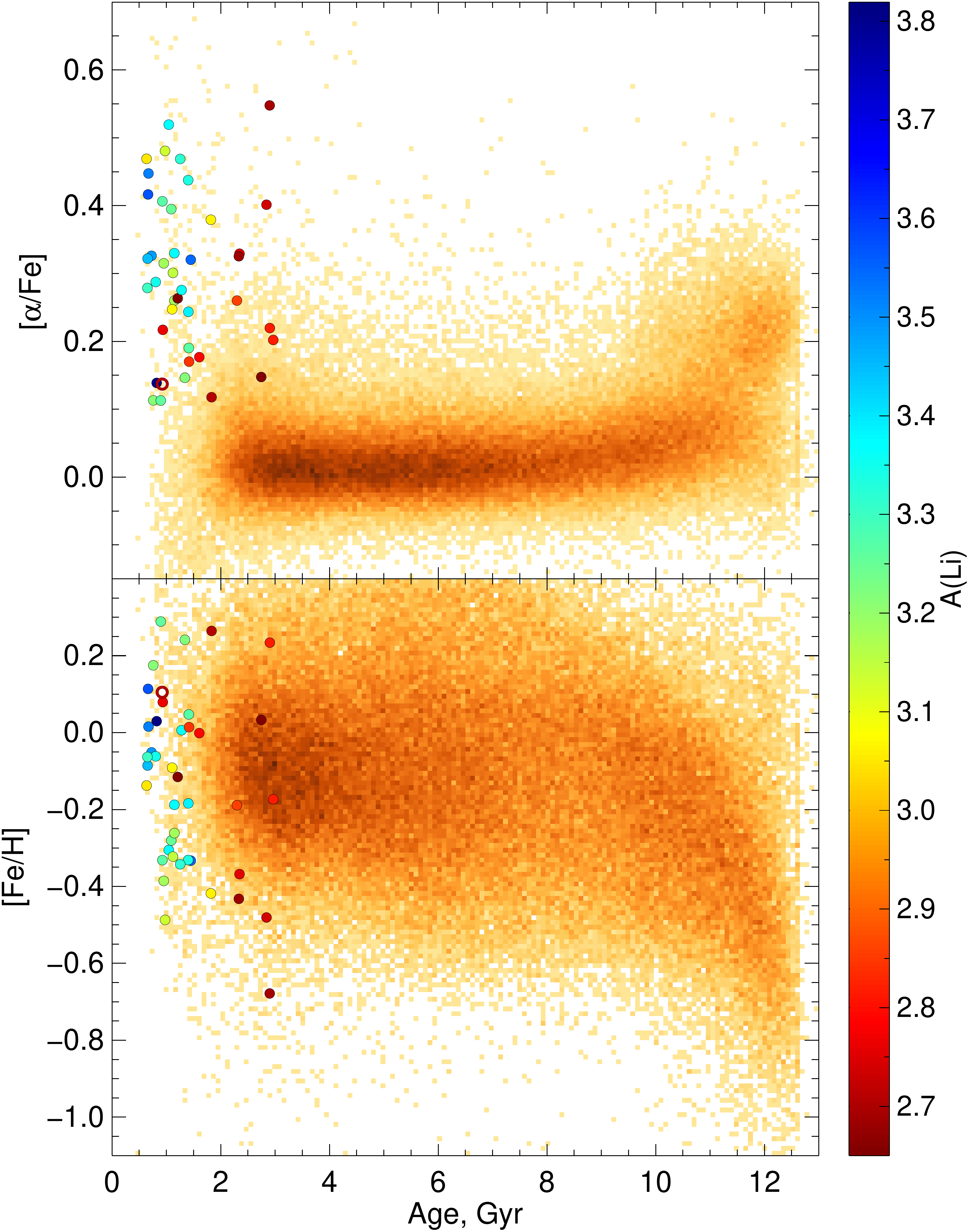}
     \caption{Number density plots showing [$\alpha$/Fe] and [Fe/H] (upper and bottom panels respectively) as a function of age for the main dwarf sample. Color-coded circles show the positions of the \exafey stars with A(Li)$>$2.65 of the main and additional samples (filled and open circles, respectively)}
     \label{afe_feh_vs_age_dwarfs}
 \end{figure}

\subsubsection{Kinematic properties}
\label{kinematic_properties}
The kinematic properties of a sample of stars also provide  interesting information on their origins. Here, we use the eccentricities and velocities of our GALAH dwarf sample, the latter being evaluated with respect to the local standard of rest ($V_{LSR}=\sqrt{\mathrm{u}^2+\mathrm{v}^2+\mathrm{w}^2}$). These values are provided in the GALAH~VAC and computed with the tool \texttt{galpy} \citep{Bovy2015}. As a star evolves, these parameters are known to increase their average values: this process of dynamical heating is caused by interactions with different substructures of the Galaxy, such as spiral arms, giant molecular clouds, or the Galactic bar \citep[e.g.,][]{Aumer2016,Mackereth2019,Almeida-Fernandes2018}.

As can be seen in Fig.~\ref{vel_ecc_distr}, our \exafey stars have significantly lower eccentricities (top panel) and velocities with respect to $V_{LSR}$ (lower panel) than old stars of the thick disk. In both cases, the distributions look very similar to those for young \anormal \ stars which are found in the thin disk. 
Thus, the kinematic properties of the \exafey stars offer further evidence in favor of characterizing them based on a young age.

\subsubsection{Discussion}
\label{DiscussionDwarfs}

Our findings in Sec.\ref{propertiesyoungalpharich} for \exafey stars can be summarized as follows: about a quarter of the \exafe stars (high \afe outliers) of our main sample have ages evaluated to less than 3~Gyr; their young age is corroborated by other independent signatures, such as the presence of stars with higher mass (hence with shorter lifetimes) than in the rest of the sample and kinematic properties similar to that of young \anormal \ stars (low eccentricities and velocities). Last but not least, one fifth of them have high Li values, which behave in the same way as those of their \anormal \ counterparts. Taken all together, the above features suggest that those \exafey stars are indeed young. It seems difficult to accept the  alternative idea of recent mergers of old, high \afe stars \citep{Zhang2021,Miglio2020}, since it would imply that such events somehow synthesize Li at the same level as in recently formed stars in the Galaxy.

In Fig.~\ref{afe_feh_vs_age_dwarfs}, we overplot our \exafey stars with high Li abundances (higher than A(Li)$_P$=2.65) on the \afe and \feh vs age diagrams of our main dwarf sample. It can be seen that the youngest of them (less than 2~Gyr) have the highest Li content and that their Li decreases with increasing stellar age; both features are compatible with what is expected for a young stellar population. The metallicity of those stars is higher than [Fe/H]=-0.4, but their high \afe ratio is the only feature that does not fit to what is expected from a young stellar population. 

That high \afe ratio could, in principle, be attributed to recent episodes of star formation in the solar neighborhood. Indeed, \citet{Mor2019} analyzed Gaia~DR2 data in combination with the Besançon Galaxy Model and found an imprint of a star formation burst $\sim$2–3~Gyr ago in the domain of the Galactic thin disk. In addition, \citet{Ruiz-Lara2020} analyzed Gaia~DR2-observed color-magnitude diagrams to obtain a detailed star formation history of the $\sim$2~kpc bubble around the Sun, which reveals three conspicuous and narrow episodes of enhanced star formation. They date those episodes as having occurred 5.7, 1.9, and 1.0~Gyr ago, also suggesting that the timing of these episodes coincides with proposed Sgr pericenter passages. 

The massive stars formed in those recent ``mini-starbursts'' eject soon after (within several Myr) their nucleosynthesis products, having an enhanced \afe ratio. If a new stellar generation is formed from those ejecta before they mix completely with the interstellar medium, stars with high \afe can be obtained having thin disk metallicities. Those formed locally or in the inner disk should have super-solar metallicities, while those in the outer disk sub-solar ones. Thus, the recent star formation episode should extend over a fairly large radial range of the thin disk in order to explain the broad range of [Fe/H] of the \exafey stars in the bottom panel of Fig.~\ref{afe_feh_vs_age_dwarfs}. In both cases, these stars could be transported to the solar neighborhood by radial migration in the thin disk, being simultaneously young, relatively massive, with thin disk kinematics and high [$\alpha$/Fe].

However, the Li content of the \exafe stars should be then depleted, since massive star ejecta are expected, in principle, to be Li poor, unless neutrino-induced nucleosynthesis manages to produce large amounts of Li during the explosion \citep{Woosley1990}. The (presently) poorly known  neutrino spectra are the main problem in evaluating the contribution of massive stars to Li production in the Galaxy. A thorough assessment of all Li sources during Galactic evolution was made in \citet{Prantzos2012}, who evaluated the maximum possible contribution of CCSN to 20\% and, more realistically, down to just a few percent. However, even in the case of such a recent star formation ``burst,'' it would be surprising to have Li at the same level as the one observed in our \exafey stars, which corresponds to standard Galactic evolution (see discussion in Sec.~\ref{high_li}). In fact, barring the case of short-lived Li sources, the Li abundance of the Galactic gas is expected to decrease in that case (although this is quantitatively difficult to estimate as it depends on many assumptions), in contrast to what is observed in the stars of our \exafey sample.

\section{Ex$\alpha$fe and \exafey red giants}
\label{high_alpha_giants}

\begin{figure*}[h!]
    \begin{multicols}{2}
    \includegraphics[width=\linewidth]{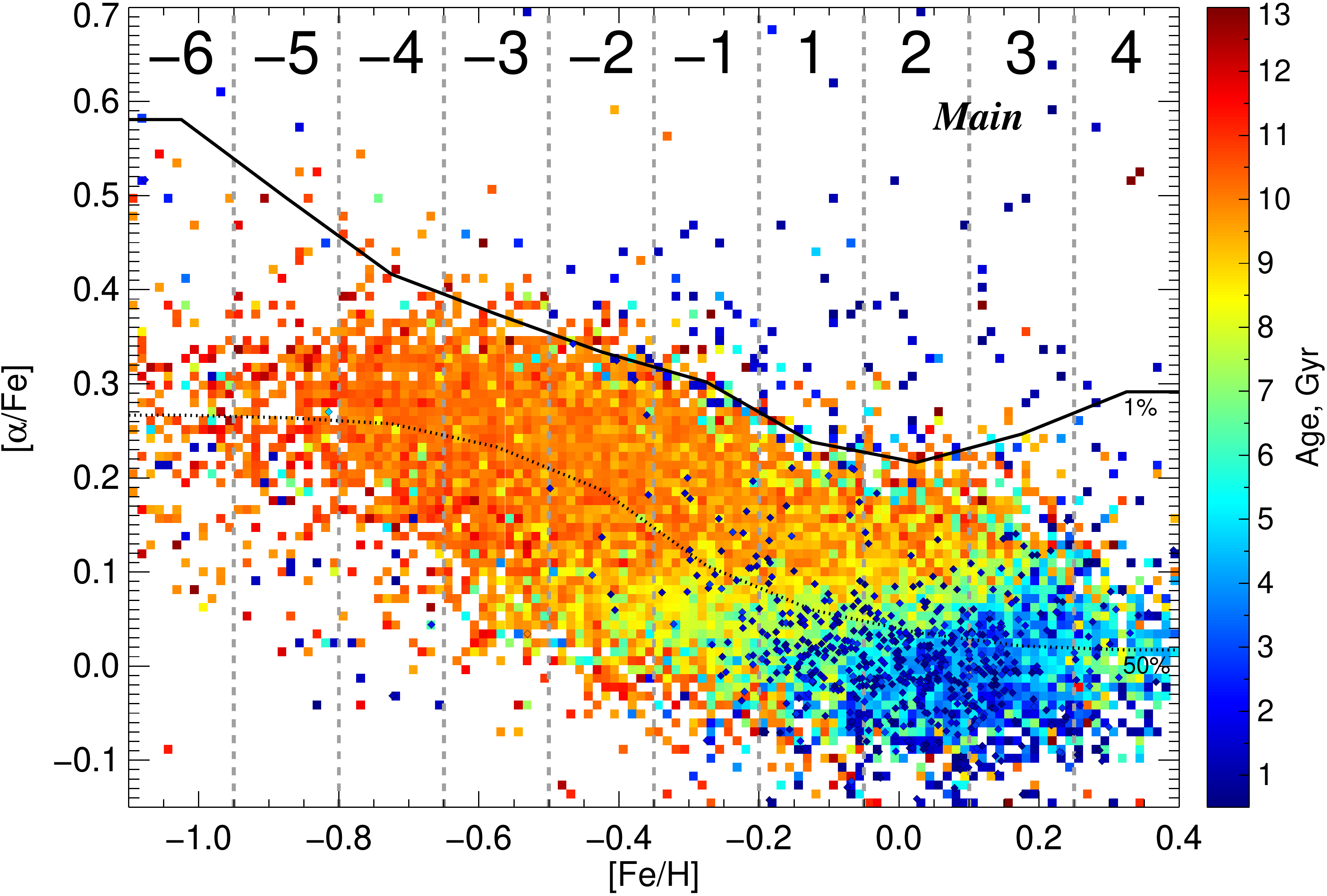}\par
    \includegraphics[width=\linewidth]{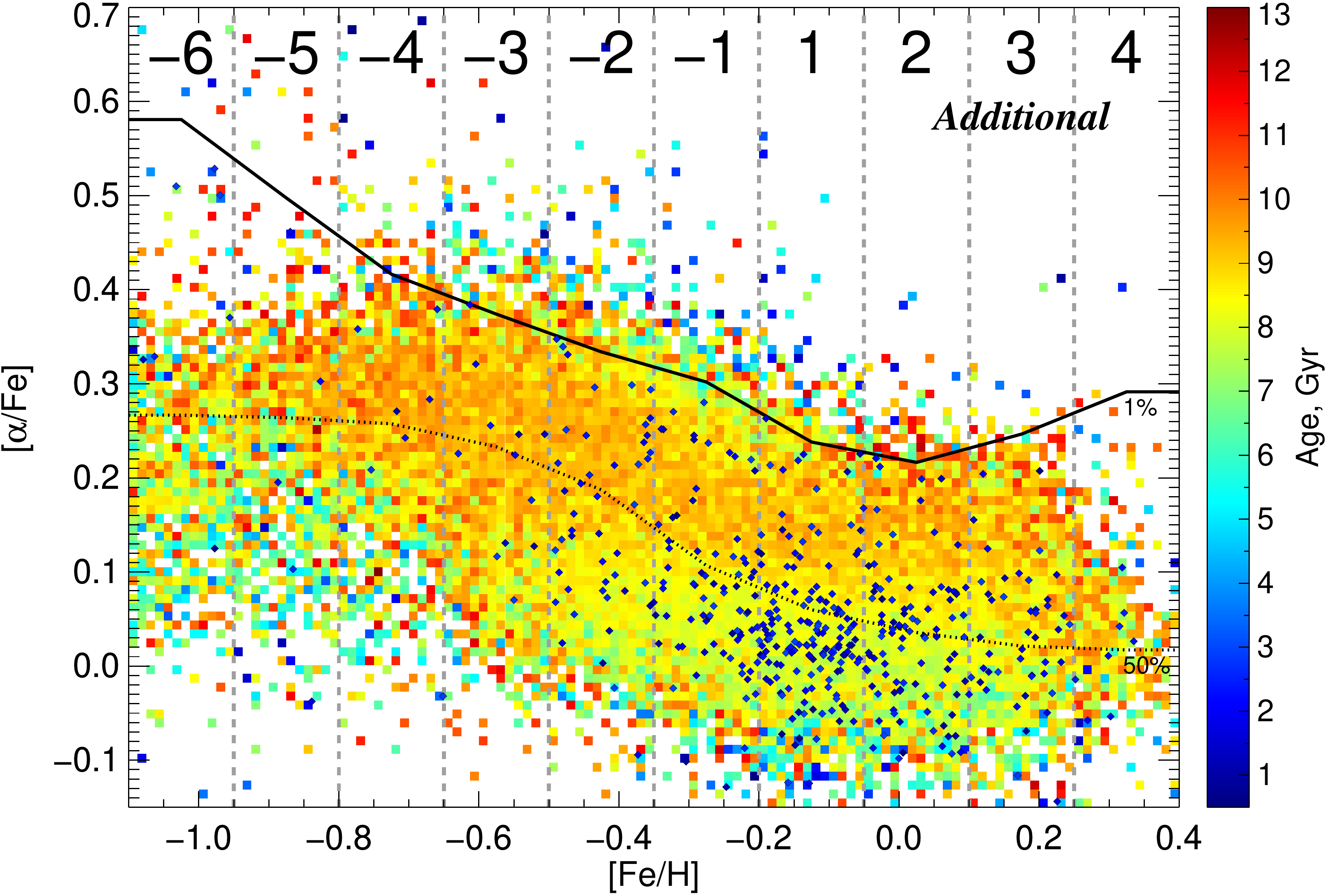}\par
    \end{multicols}
    
    \begin{multicols}{2}
    \includegraphics[width=\linewidth]{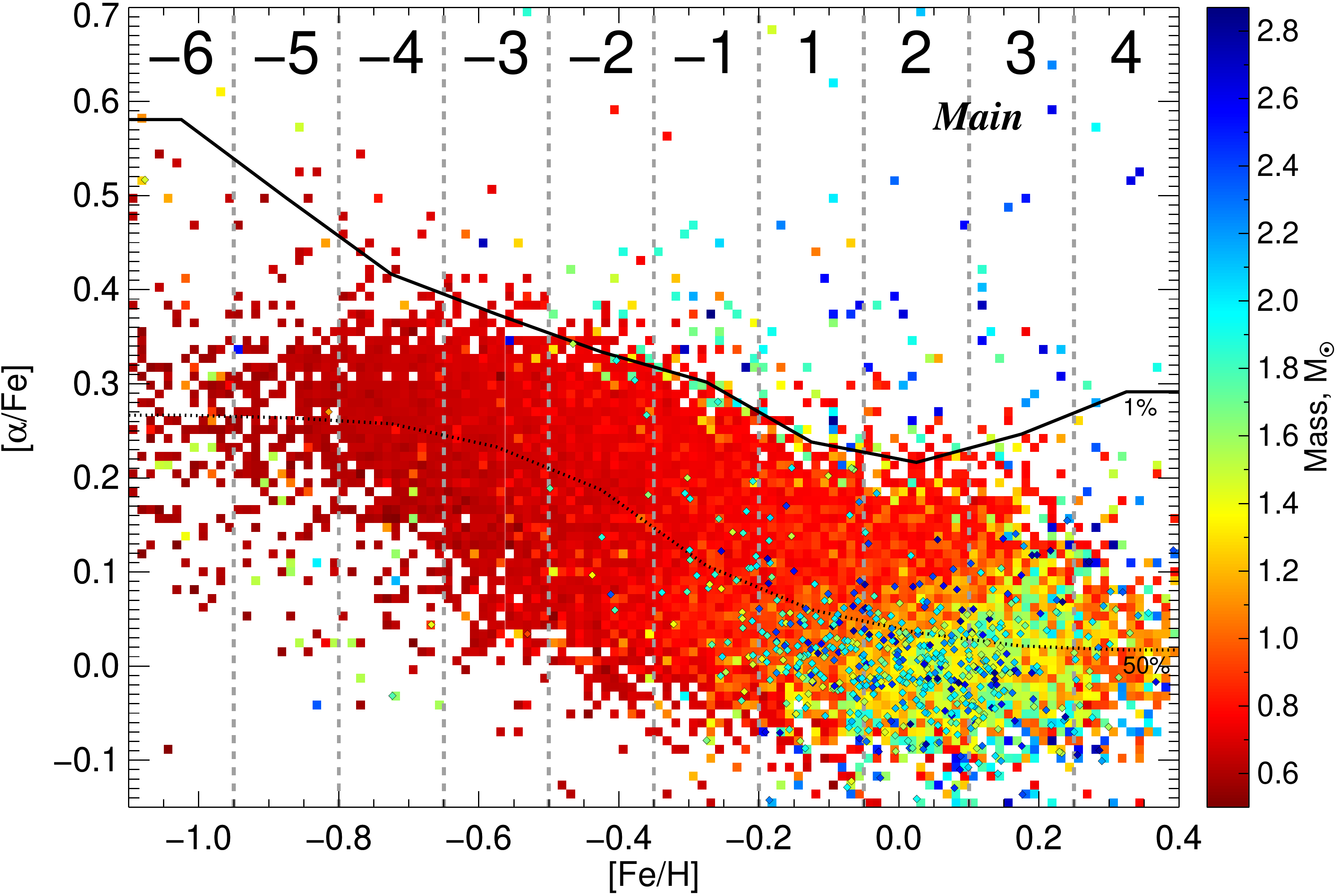}\par
    \includegraphics[width=\linewidth]{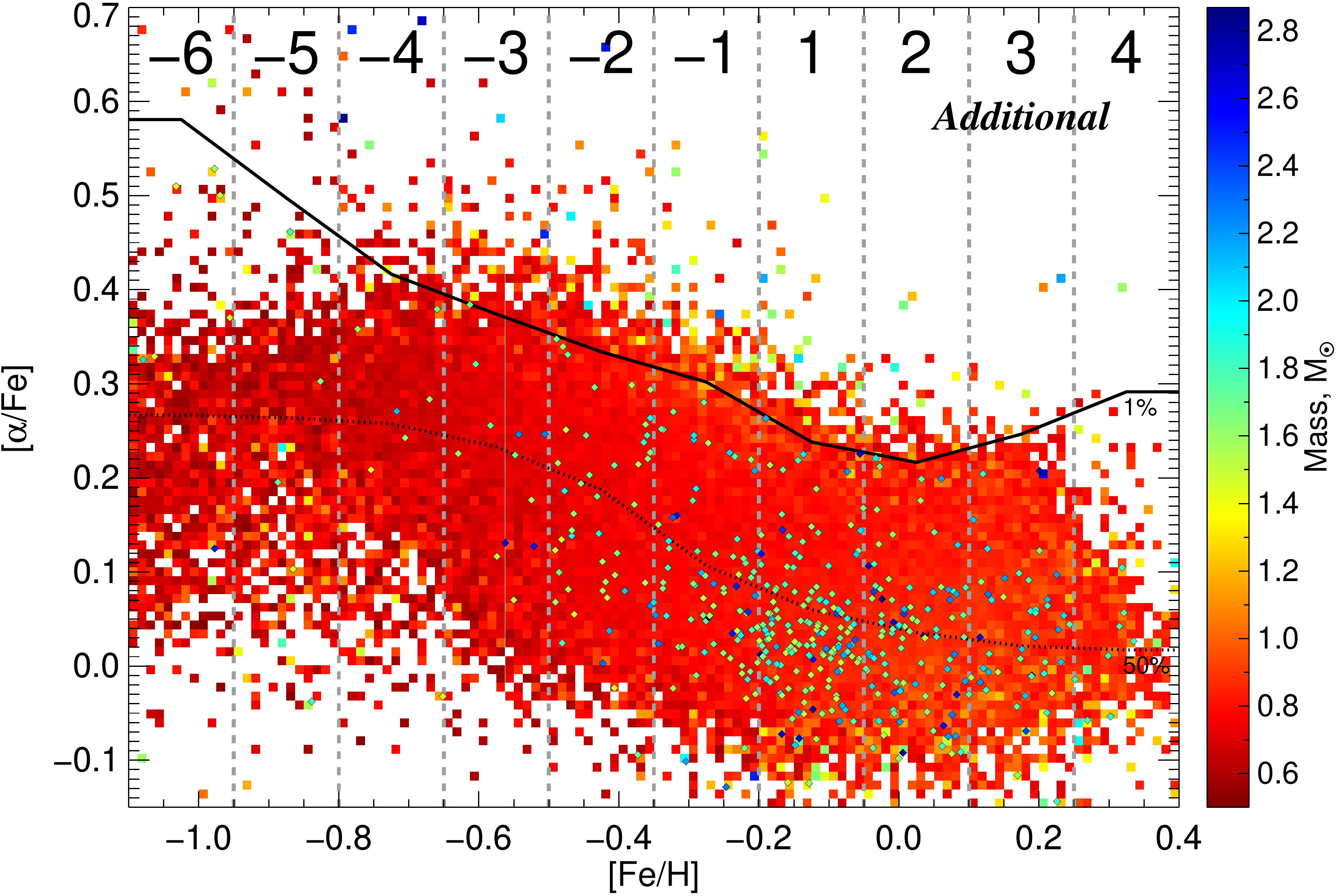}\par
    \end{multicols}
         \caption{Properties of the giants of the main and additional samples (left and right columns, respectively) in the Tinsley-Wallerstein diagram. Small squares represent bins with more than five stars. Blue points show the positions of young ($<$3~Gyr) stars.}
     \label{afe_vs_feh_age_giants}
 \end{figure*}
 
 We now turn to \exafe and \exafey red giants. We applied the same selection criteria as for the dwarfs presented above and so, we discuss the implications of these differences on the conclusions of previous studies of so-called young [$\alpha$/Fe]-high red giants.
 
\subsection{Sample selection} 
\label{sample_selection_giants}
We selected giant stars from GALAH~DR3 applying the same criteria as for the dwarf main and additional samples, except for the log~$g$ and $T_\mathrm{eff}$ domains (Fig.~\ref{kiel_diagram}). The age determination through the isochrone-based method leads to higher uncertainties for giant than for dwarf stars, especially in the low-mass range ($\leq$2~M$_{\odot}$) where evolution tracks are close from the Hayashi limit up to the RGB tip. As a result, we have about three times more giant stars in the additional sample than in the main one (we had the inverse ratio for the dwarfs), with a large fraction in the low-mass range. We identified the \exafe giants as those above the $1~\%$ curve that is re-computed for the entire giant sample (Fig.~\ref{afe_vs_feh_age_giants}). Finally, we did not consider Li as an age indicator since its abundance is strongly affected in red giants during the first dredge-up episode and later above the RGB bump \citep[][and references therein]{2020A&A...633A..34C}. The information about the resulting main and additional samples is summarized in Table~\ref{tab_sample_giants}, and Fig.~\ref{kiel_diagram} shows the positions of the selected giants in the Kiel diagram. 

\begin{table*}[h!]
\centering
\begin{tabular}{|l|ccc|ccc|} 
\hline
& Main & \exafe &\exafey & Additional & add-\exafe & add-\exafey \\
Criterion          &   &   &   & &   &  \\
\hline
\hline
log~$g$ & $<$3.8 & $<$3.8 & $<$3.8 & $<$3.8 & $<$3.8 & $<$3.8 \\\
$T_\mathrm{eff}$ & 3000$\div$5500 & 3000$\div$5500 & 3000$\div$5500 & 3000$\div$5500 & 3000$\div$5500 & 3000$\div$5500 \\\
[Fe/H]          &   -1.1$\div$0.4~dex   & -1.1$\div$0.4~dex & -1.1$\div$0.4~dex & -1.1$\div$0.4~dex & -1.1$\div$0.4~dex & -1.1$\div$0.4~dex\\\
[$\alpha$/Fe]   &   -0.15$\div$0.7~dex  & top 1\%  of Main         &   top 1\% of Main        & -0.15$\div$0.7~dex &  top 1\% of Add  &  top 1\% of Add \\
$\sigma_{[\alpha/\mathrm{Fe}]}$   & $\leq0.05$~dex  & $\leq0.05$~dex & $\leq0.05$~dex & $\leq0.05$~dex & $\leq0.05$~dex & $\leq0.05$~dex\\
Age &   0.5$\div$13~Gyr &   0.5$\div$13~Gyr & 0.5$\div$3~Gyr  & 0.5$\div$3~Gyr & 0.5$\div$3~Gyr & 0.5$\div$3~Gyr\\
$\sigma_\mathrm{age}/\mathrm{age}$ & $\leq$30\%    & $\leq$30\% & $\leq$30\% & $>$30\% & $>$30\% & $>$30\% \\
S/N & $\geq$30   & $\geq$30 & $\geq$30    & $\geq$30 & $\geq$30 & $\geq$30 \\
{\small{\texttt{flag\_sp}}} & +    & + & + & + & + & + \\
{\small{\texttt{flag\_fe\_h}}} & +    & + & +  & + & + & +\\
{\small{\texttt{flag\_alpha\_fe}}} & +    & + & +  & + & + & +\\
\hline
\hline
Total number & 23~072       &  235      &  107    & 83~213    & 990   & 93  \\
\hline
\end{tabular}
\caption{Selection criteria applied to build the different sub-samples of giants from the GALAH catalog.} 
\label{tab_sample_giants}
\end{table*}
 
 \begin{figure}[h!]
    \includegraphics[width=\linewidth]{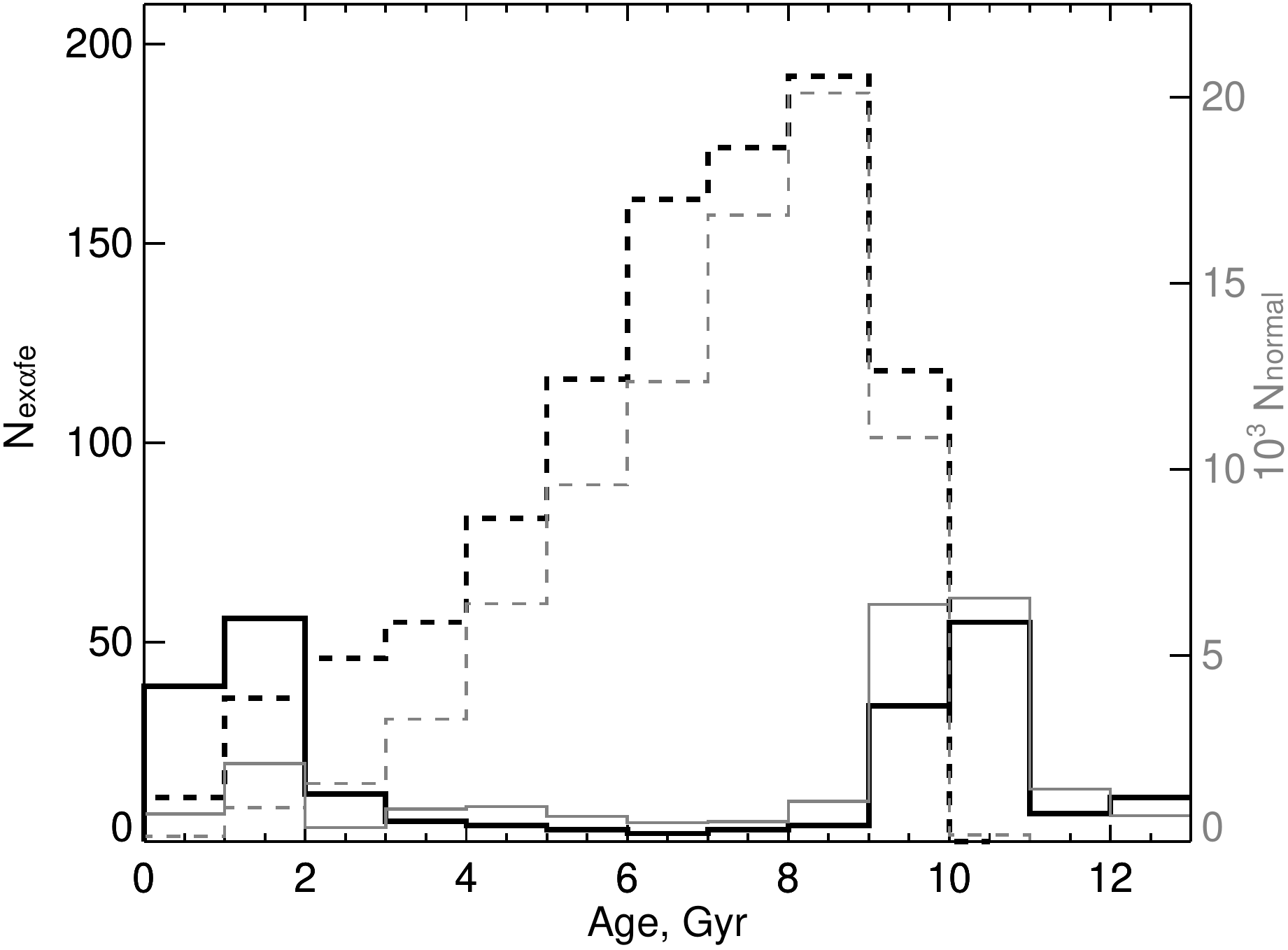}
    \includegraphics[width=\linewidth]{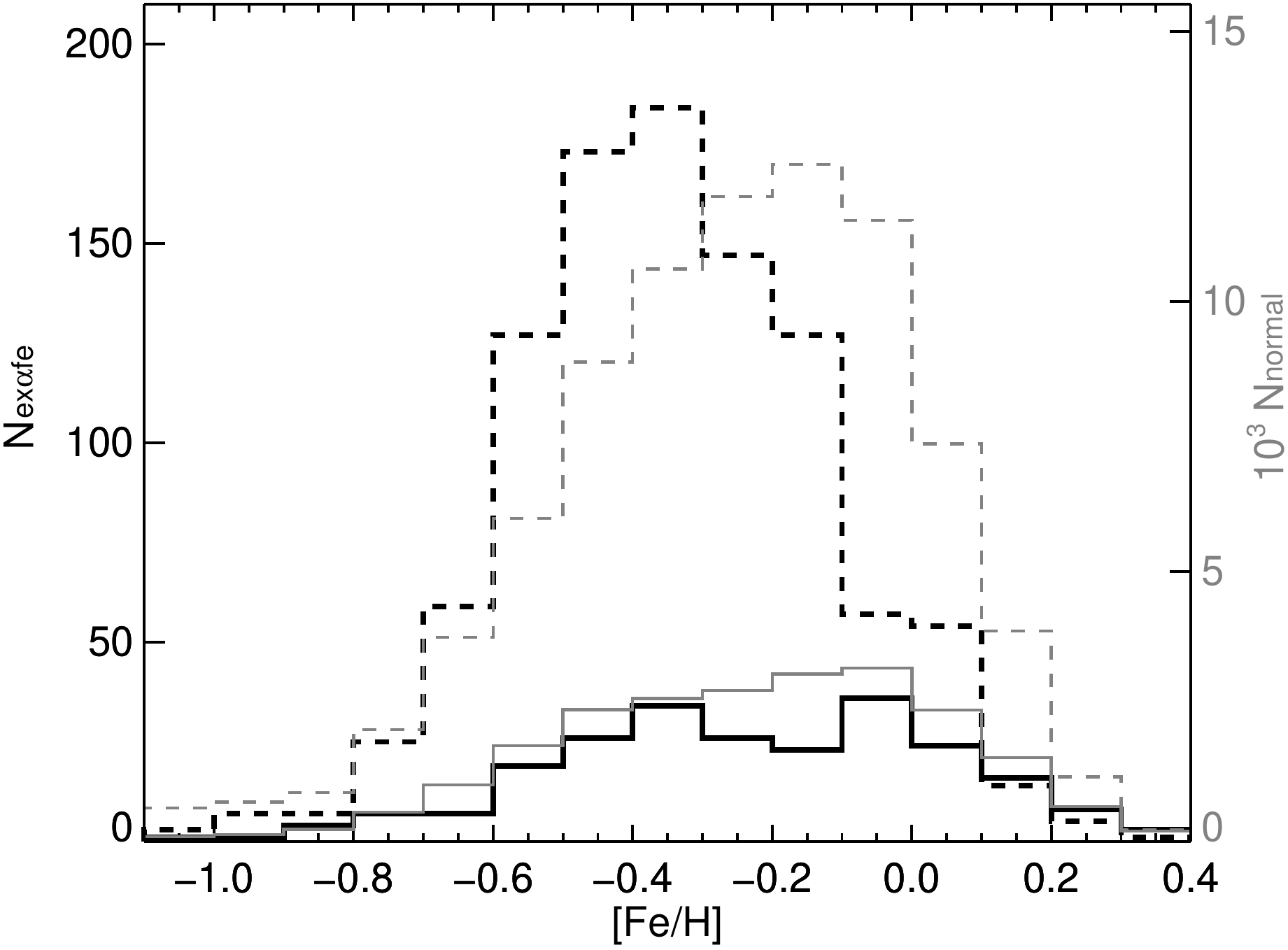}
     \caption{Age ({\it{top}}) and [Fe/H] ({\it{bottom}}) distributions of the \exafe (thick black, scale on the left) and \anormal \ (thin grey, scale on the right) giant stars from the main (solid) and additional (dashed) samples.} 
     \label{hist_age_feh_alpha_rich_giants}
 \end{figure}

 \begin{figure}[h!]
    \includegraphics[width=\linewidth]{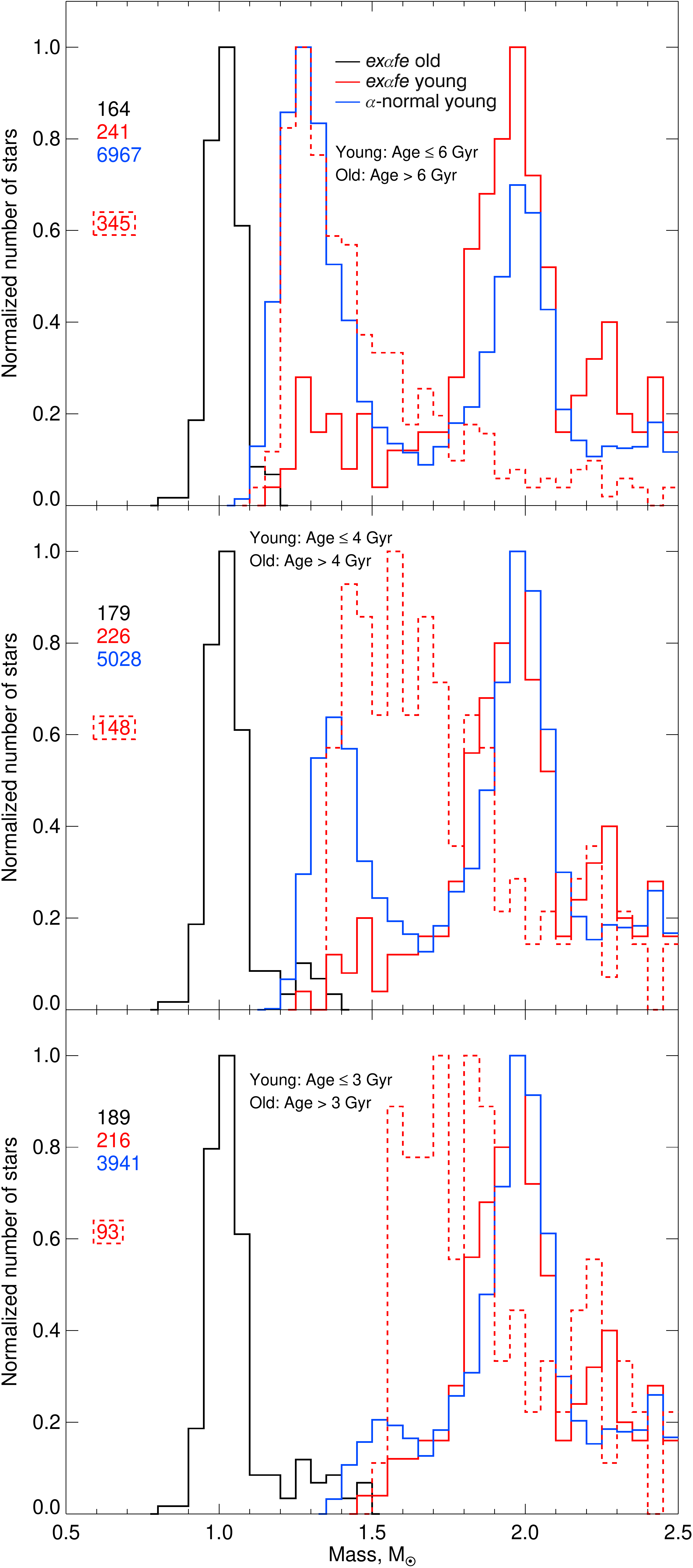}
     \caption{Mass distribution of the giant stars assuming different age limits (3, 4, and 6~Gyr from bottom to top). Red histograms correspond to \exafey (i.e., younger than the age cut) giants of the main and additional samples (full and dotted respectively). The blue and black histograms correspond respectively to the young \anormal \ giants and to \exafe giants older than the cut (for these two cases we show only the distributions of the main sample stars). All the histograms are scaled to have the same height for the purpose of visibility, and the numbers on the left give the number of stars in each color-coded class of giants.} 
     \label{mass_distribution_lamost_criteria_giants_new}
 \end{figure}

\subsection{Properties of the \exafe and \exafey giants}
\label{exafe_exafey_giants}
The ages and masses of the selected giant stars 
are color-coded in the Tinsley-Wallerstein diagram in Fig.~\ref{afe_vs_feh_age_giants}, and we show the age and [Fe/H] distributions of the \exafe and \anormal \ stars in Fig.~\ref{hist_age_feh_alpha_rich_giants} (compare respectively to Fig.~\ref{afe_vs_feh_age_mass_Li_main_add} and \ref{hist_age_feh_alpha_rich} for the dwarfs). As in the case of the dwarfs, the [Fe/H] distributions of the \exafe and \anormal \ giants are very similar, and we retrieve the global decrease of \afe with decreasing stellar age for the \anormal \ giants. The age distribution of the \exafe giants from the main sample is similar to that of the \exafe dwarfs, albeit with a larger fraction being younger than 3~Gyr (compare to Fig.~\ref{hist_age_feh_alpha_rich}). The additional sample also contains \exafey giants, although it is dominated by old low-mass stars whose evolution tracks along the RGB make the age determination more uncertain. Finally, \exafey giants are found in all metallicity bins, both in the main and additional samples, as observed in the dwarf sample.

We show in Fig.~\ref{mass_distribution_lamost_criteria_giants_new}, the mass distributions (normalized numbers) of the old {{ex$\alpha$fe}}, y-ex$\alpha$fe, and young \anormal \ giants of the main sample; regarding the additional sample, we only show the mass distribution of the \exafey stars, for  the purpose of clarity. We separated young and old stars assuming different age cuts, namely, 3 (as for the dwarfs), 4, and 6~Gyr, for a comparison with previous works (\S~\ref{comparison_previous_work_giants}). For the 3~Gyr case, two peaks appear in the mass distribution of the main sample. On one hand, the peak located around $\sim$ 0.9-1.0~M$_{\odot}$ corresponds to the \exafe giant stars with age$>$3~Gyr that are climbing the RGB and whose age distributions peak around 10~Gyr (Fig.~\ref{hist_age_feh_alpha_rich_giants}), which is approximately the main sequence lifetime of 0.9-1.0~M$_{\odot}$ stars at [Fe/H]$\sim$-0.5~dex. This explains why its position does not move when we increase the age cut. On the other hand, the mass distributions of the \exafey and young \anormal \ giants nicely overlap at 3~Gyr, with a peak around 1.9-2.0~M$_{\odot}$ which contains stars that are presently close to or at the red clump (see Fig.~\ref{kiel_diagram}). When we increase the age cut, this peak slightly moves towards lower masses, due to the dependence of the stellar lifetime with mass. An additional peak appears in the distribution of the young \anormal \ stars at intermediate masses (at $\sim$ 1.4 and 1.3~M$_{\odot}$ for the age cuts of 4 and 6~Gyr, respectively, and barely visible at 1.5~M$_{\odot}$ at 3~Gyr). It corresponds to stars whose main sequence lifetime is similar to or lower than the age cut, and which are climbing the RGB. Its position in mass thus decreases with increasing age limit. Because the age determination for giants in the corresponding mass domain is rather uncertain, this peak is much more pronounced for the \exafey from the additional than for those of the main sample (compare red full and dotted lines). Actually, when we consider the main and the additional samples together, the relative height of the RGB peak is much higher than that of the clump stars, and approximately at the same position in mass for the \exafey and the young \anormal \ giants. 

Kinematics of giants (both eccentricities and V$_{LSR}$) also tends to the fact that \exafey giants are indeed young. However, distributions of the kinematic parameters of \exafey and young \anormal \ are not absolutely similar, as the former exhibit lower eccentricities and V$_{LSR}$ than old {{ex$\alpha$fe}}.

  \begin{figure}[h!]
   \center
     \includegraphics[width=\linewidth]{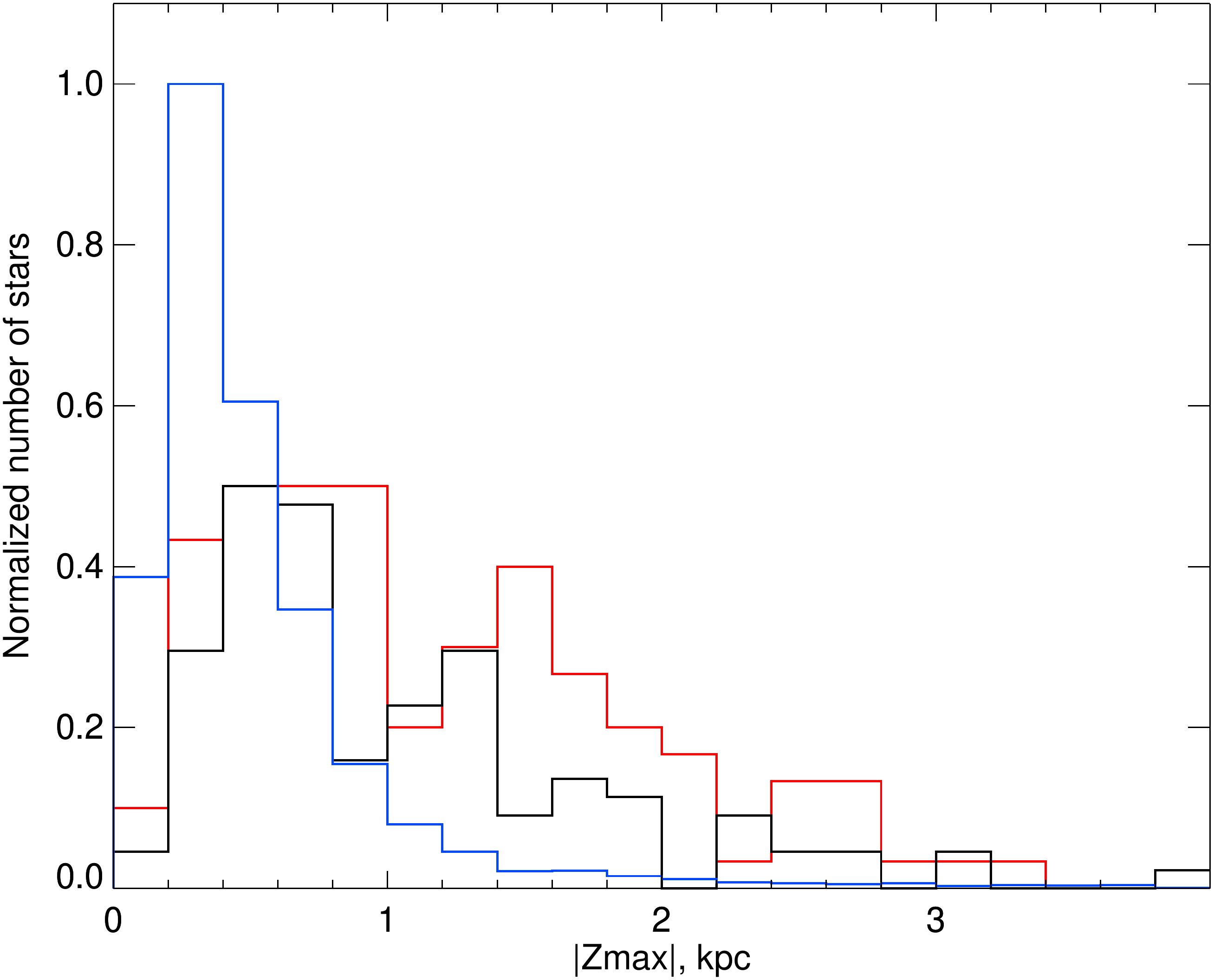}
     \includegraphics[width=\linewidth]{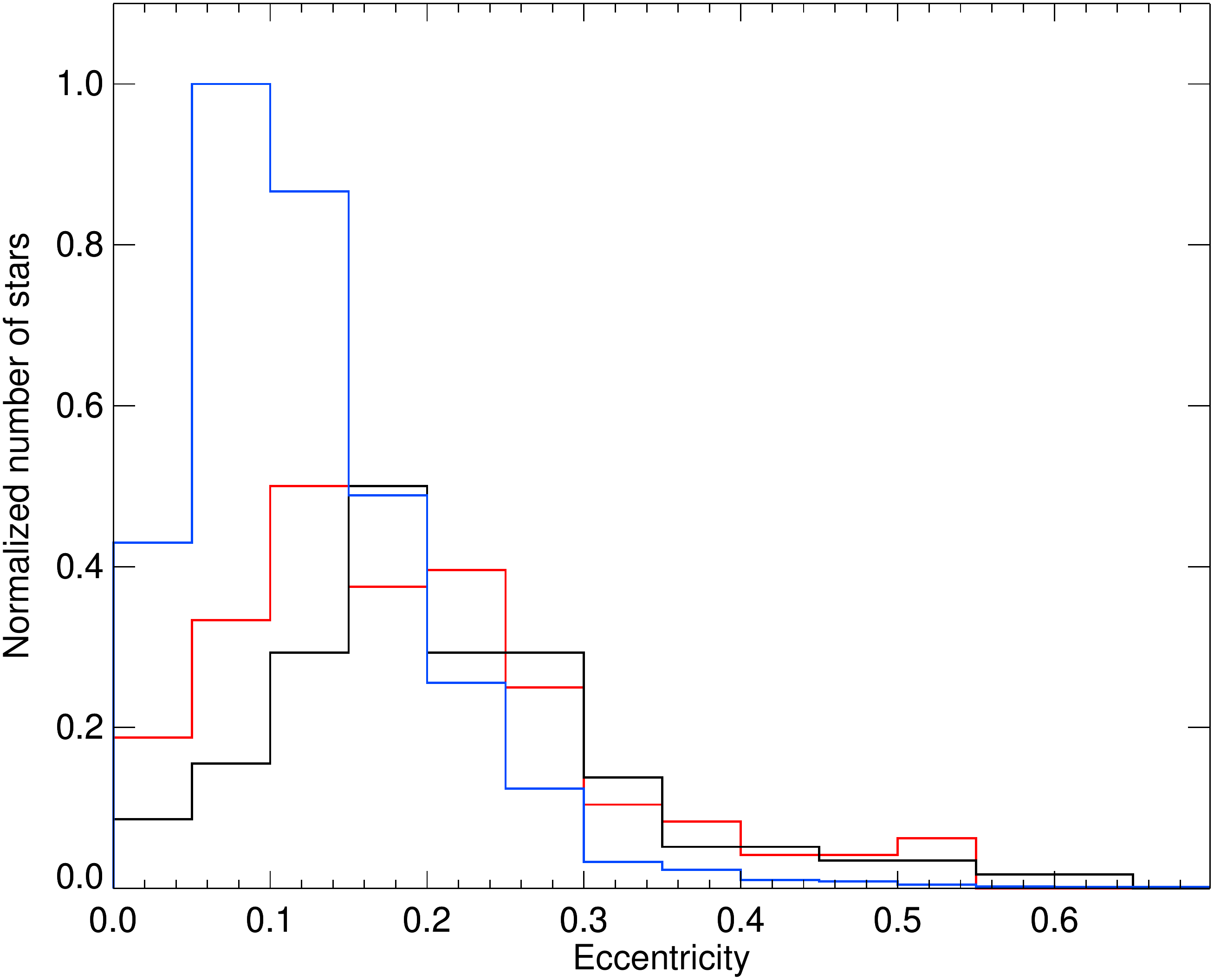}
     \includegraphics[width=\linewidth]{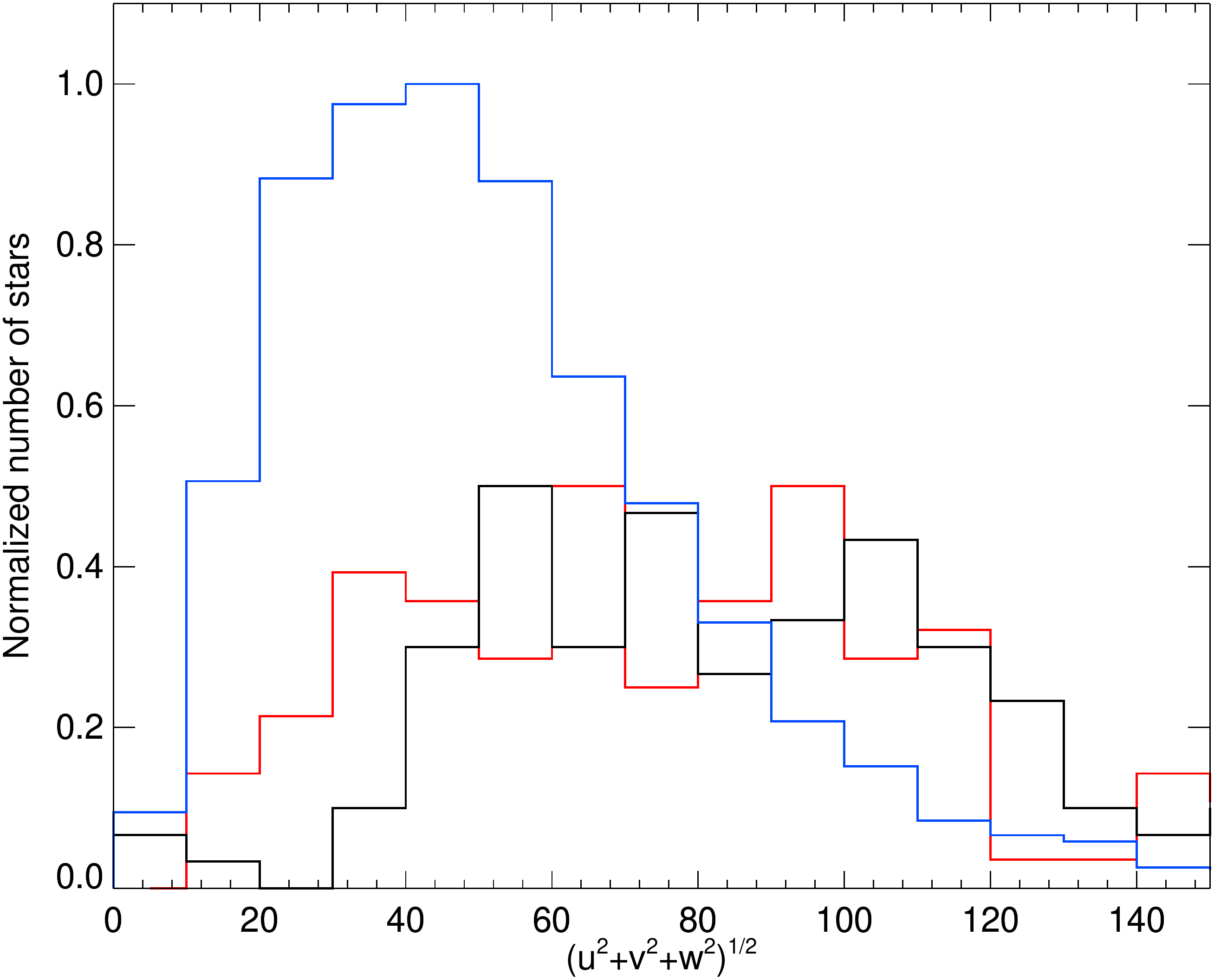}
     \caption{Distribution of the maximum Galactocentric orbit height Z$_{max}$ (upper panel), eccentricity (middle panel), and V$_{LSR}$ of the giants (same colors as in Fig.~\ref{mass_distribution_lamost_criteria_giants_new}).}
     \label{kinematics_giants_exafe}
 \end{figure}

We conclude from this analysis that: 1) the mass and metallicity distributions are the same for the \exafey and the young \anormal \ giants; 2) the positions of the peaks in masses found above for the giants stars reflect secular evolution effects, and do not result from an odd IMF; 3) their apparent respective heights depend on the criteria we impose on the precision of the age determination, but the ``RGB peak'' dominates over ``clump peak'' when we combine the main and the additional samples; 4) the kinematic parameters of \exafey giants and young \anormal \ giant stars are very similar, and they differ from those of \exafe and \anormal \ giants with age$>$3~Gyr. This reinforces the conclusions we draw for the dwarf stars that the \exafey stars did not form through mergers but, rather, with an IMF similar to that of their \anormal \ counterparts.

\subsection{Comparison to previous work}
\label{comparison_previous_work_giants}

Previous studies of so-called ``young, high [$\alpha$/Fe]'' red giant stars used different criteria both in terms of [$\alpha$/Fe] and age, with 6~Gyr being usually assumed as the transition between young and old populations. 
The discovery of such stars in the Galactic disk in CoRoT-APOGEE (CoRoGEE) and Kepler-APOGEE (APOKASK) samples gave a new spin to the field, as asteroseismology has opened a new avenue to determine the age and mass of giants and to distinguish clump stars from RGB stars \citep[e.g.,][]{Prantzos2012,Chiappini2015,Martig2015,Anders2017,2018MNRAS.475.5487S}. 

\begin{figure*}[h!]
   \center
     \includegraphics[width=0.45\linewidth]{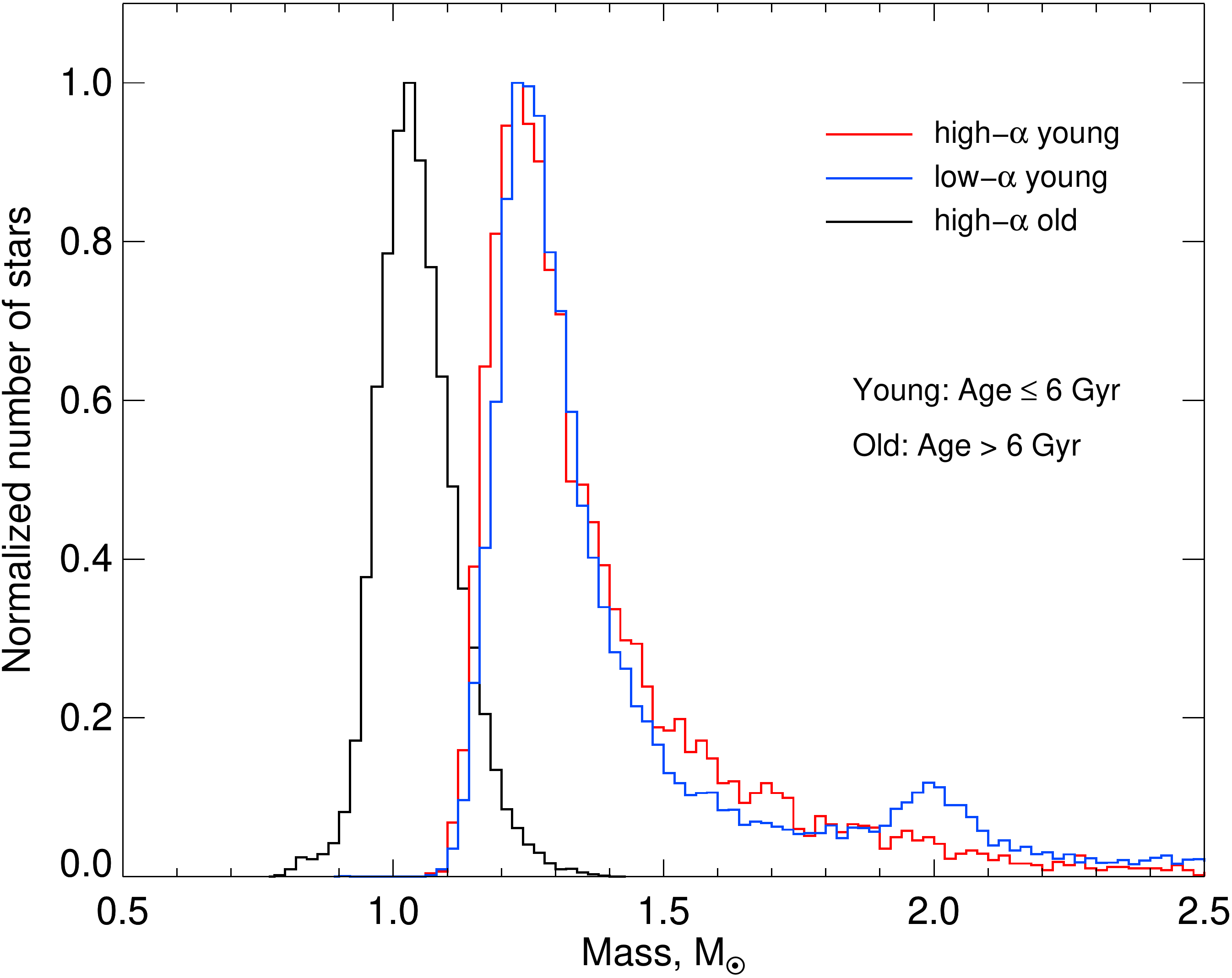}
     \includegraphics[width=0.45\linewidth]{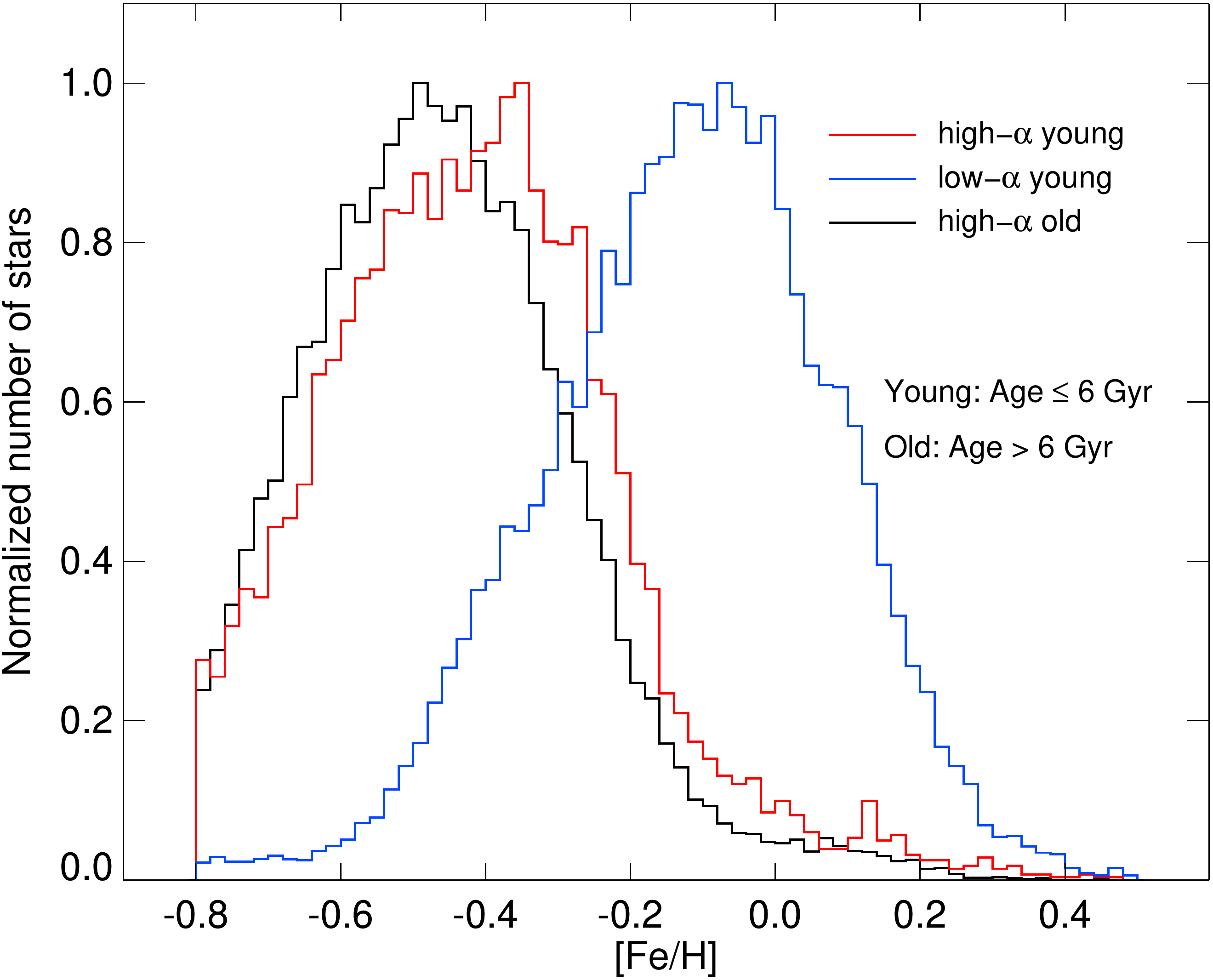}
     \caption{Mass and [Fe/H] distribution (left and right panels, respectively) of the GALAH~DR3 giants selected with the same criteria on parameters as in \citet{Zhang2021}. All the histograms are scaled to have the same height for the purpose of better visibility.}
     \label{mass_distribution_lamost_criteria_giants}
 \end{figure*}

\citet{Zhang2021} expanded the statistics using LAMOST~DR4 value-added catalog \citep{2019MNRAS.484.5315W,2019ApJS..245...34X}, with ages and masses based on data-driven spectroscopic estimates ($T_\mathrm{eff}$, log~g, C and N abundances) trained by a Kepler asteroseismic sample. They eliminated red clump stars and kept only RGB stars with 3000~$<$~$T_\mathrm{eff}<$~5500~K, log~$g<3.8$, and $-0.8<$[Fe/H]$<0.5$~dex. They defined ``high-$\alpha$ young giant stars'' those with \afe higher than 0.18~dex independently of their [Fe/H] value, and with ages lower than 6~Gyr; they did not exclude stars based on age uncertainties. With their selection criteria, they obtained the peaks around 0.9-1.0~M$_{\odot}$ for the ``high-$\alpha$ old'' RGB stars and at $\sim$ 1.2~M$_{\odot}$ for the ``high-$\alpha$ young'' RGB stars, as in our sample when we consider 6~Gyr for the age cut. Obviously, the peak we find in our sample around 1.9-2.0~M$_{\odot}$ could not appear in the mass distribution of their RGB sample, because it corresponds to clump stars that were excluded from their analysis. As far as the masses of the red giants are concerned, the two studies thus agree nicely. However, the bulk of their ``low-$\alpha$'' young stars appear to be more metal-rich (by about 0.3~dex) than the distributions of their ``high-$\alpha$'' RGBs, young and old. To check if this difference in the [Fe/H] behavior results from the different limits in [$\alpha$/Fe] we use, we repeated the analysis on GALAH~DR3 using \citet{Zhang2021} criteria. The corresponding sample contains 118~464 giants, out of which 35~957 have \afe higher than 0.18~dex. As shown in Fig.~\ref{mass_distribution_lamost_criteria_giants} for the age limit of 6~Gyr (here, we do not split the set into main and additional samples so that we can  maintain consistency with \citealt{Zhang2021}), we retrieve the same patterns as in their study (compare to their Fig.~5). Since we have no criteria to discriminate clump stars from the other giants in GALAH, the only difference in the mass distribution is the corresponding peak appears at $\sim$2~M$_{\odot}$ for the ``low-$\alpha$ young giant stars'', with a relative height much smaller than that of the RGB peak, as expected for a normal IMF. As for the [Fe/H] distribution, we retrieve the shift they obtained between the ``low-$\alpha$ young giant stars'' on one hand, and the young and old ``high-$\alpha$'' RGBs on the other hand. This shift is thus simply due to the \afe limit adopted to distinguish between $\alpha$-made in \citet{Prantzos2012}, who evaluated the maximum possible enriched and $\alpha$-normal stars. In addition, we retrieved similar distributions of other elements presented in their Fig.~8 (except for nitrogen which is absent in the GALAH~DR3 data).

To conclude, the results we draw based on GALAH~DR3 data are fully consistent with those obtained by \citet{Zhang2021} using LAMOST for giants when we use the same selection criteria for young $\alpha$-enriched stars. However, while \citet{Zhang2021} call for stellar mergers to explain the mass distribution of their ``high-$\alpha$ young giant stars,'' based on the fact that 1.2~M$_{\odot}$ is significantly higher than typical masses of thick disk stars (1~M$_{\odot}$), our study depicts the effects of secular evolution for a population of \exafey giants born with a normal IMF and a [Fe/H] distribution similar to that of their \anormal \ counterparts. This is in agreement with our conclusions about the \exafey dwarf stars.

\section{Summary and conclusions}
\label{sec:summary}

In this paper, we focus, for the first time on dwarf and red giant stars with ``extreme'' high \afe values (the upper 1$\%$, independently of their [Fe/H], with [Fe/H] between -1.1 and +0.4~dex.), which we call {{ex$\alpha$fe}}, and on the youngest of them (ages below 3~Gyr), which we refer to as ``y-ex$\alpha$fe''. We selected our sample from the GALAH~DR3 survey, applying strict criteria on the precision of the stellar parameters, abundances, and age and mass determinations, and we excluded binaries. We identified a large fraction of \exafey stars among both the dwarf and giant \exafe populations ($\sim 30 \%$ and $\sim 15 \%,$ respectively) and showed that the [Fe/H] distribution of the {{ex$\alpha$fe}}, y-ex$\alpha$fe, and \anormal \ stars overlap. Beyond the isochrone-based age determination method for single stars provided by GALAH~DR3 that we successfully compared to the ages we determined with a different Bayesian tool on another set of stellar models, we explored other indicators of ages, namely, Li abundances for the dwarfs and kinematics for both the dwarf and red giant samples. We also investigated the mass distribution of the overall sample.

As for the \exafey dwarf stars, both the Li and the kinematics support their young age. First, the youngest (according to the classical age determination method), hottest, and most massive ones exhibit Li abundances slightly higher (by a factor of 2) than the proto-solar value, in agreement with the Li observed in their young  $\alpha$-normal counterparts. This means that the two populations were born with essentially the same initial Li abundances. Additionally, we find that the \exafe and the $\alpha$-normal dwarfs undergo the same mass-dependent Li depletion with time, which is also observed in open clusters. Second, the kinematic properties of the \exafey dwarf stars (lower eccentricities of their orbits and lower $V_{LSR}$ compare to those of old stars of the thick disk) are similar to those of young \anormal \ stars, providing strong evidence that they belong to the young stellar population of the thin disk. Finally, after accounting for secular evolution effects and for the magnitude limitations of the GALAH survey that induce artificial patterns in the mass distribution of the entire sample, we showed that \exafey dwarfs do not have higher average mass compared to \exafe and \anormal \ stars.

We also considered \exafe and \exafey giants from GALAH, selecting them with the same criteria as for the dwarfs. We did not use the Li indicator, since the first dredge-up induces strong surface Li depletion when the stars become giants, blurring out the information on the initial Li abundance the stars were born with. As for the other properties, our analysis leads to the same conclusions as for the \exafey dwarfs. Namely, \exafey giants are indeed young both in terms of isochrone-based age determination and kinematics indicators. In addition, they have the same [Fe/H] distribution as young \anormal \ giant stars. Last but not least, we showed that the position of the peaks in the mass distribution of both the \exafey and young \anormal \ giants results from secular evolution. In particular, the peak at $\sim$ 1.9-2~M$_{\odot}$ we obtain with the adopted age limit of 3~Gyr corresponds to stars at or near the clump. When we increase the age cut, this peak naturally moves towards lower stellar masses that have longer lifetimes, and the RGB peak at lower mass becomes  dominant as lower mass stars make it to the red giant phase.   

We investigated the impact of the different selection criteria and age limits used in literature studies of giant stars with high [$\alpha$/Fe]. We showed in particular that the mass and [Fe/H] distributions of the ``high-$\alpha$ young giant stars'' that lead \citet{Zhang2021} to support the merger scenario is due to a combination of stellar secular evolution and of the adopted limits for both \afe and the ages of their sample. This strengthens our conclusions that \exafey dwarf and giant stars did not form through mergers, but rather with a similar IMF than \anormal \ stars.

The origins of the \exafey stars, which compose $\sim$0.3 \% of the large sample of dwarf and red giant stars of the local Galactic population that we selected with strict quality criteria from GALAH~DR3, is still unclear. Further studies are needed to check whether their high $[\alpha$/Fe] ratio reflects massive star ejecta in recent enhanced star formation episodes in the Galactic thin disk, resulting, for instance, from interactions with the Sagittarius dwarf galaxy.

 \begin{acknowledgements}
 This work was supported by the Swiss National Science Foundation (Project 200020-192039 PI C.C.). We extensively used NASA's Astrophysics Data System Bibliographic Services and TOPCAT \citep{Taylor2005}. We are grateful to A.~De~Cia, N.~Lagarde, V.~Hill, and E.~Fern\'{a}ndez Alvar for enlightening discussions. We are thankful to the anonymous referee for the important comments.
\end{acknowledgements}

\bibliographystyle{aa}
\bibliography{Alpha-rich_stars}

\end{document}